\documentclass[sn-mathphys,Numbered]{sn-jnl}

\usepackage{graphicx}%
\usepackage{multirow}%
\usepackage{amsmath,amssymb,amsfonts}%
\usepackage{amsthm}%
\usepackage{mathrsfs}%
\usepackage[title]{appendix}%
\usepackage{xcolor}%
\usepackage{textcomp}%
\usepackage{manyfoot}%
\usepackage{booktabs}%
\usepackage{algorithm}%
\usepackage{algorithmicx}%
\usepackage{algpseudocode}%
\usepackage{listings}%

\theoremstyle{thmstyleone}%
\theoremstyle{thmstyletwo}%

\theoremstyle{thmstylethree}%

\begin{document}

\title[Remnants of quark model in lattice QCD simulation in the Coulomb gauge]{Remnants of quark model in lattice QCD simulation in the Coulomb gauge}


\author*[1]{\fnm{Hiroki} \sur{Ohata}}\email{hiroki.ohata@yukawa.kyoto-u.ac.jp}

\author[2]{\fnm{Hideo} \sur{Suganuma}}\email{suganuma@scphys.kyoto-u.ac.jp}
\equalcont{These authors contributed equally to this work.}

\affil*[1]{\orgdiv{Yukawa Institute for Theoretical Physics}, \orgname{Kyoto University}, \orgaddress{\street{Kitashirakawa oiwake, Sakyo}, \city{Kyoto}, \postcode{606-8502}, \state{Kyoto}, \country{Japan}}}

\affil[2]{\orgdiv{Department of Physics, Graduate School of Science}, \orgname{Kyoto University}, \orgaddress{\street{Kitashirakawa oiwake, Sakyo}, \city{Kyoto}, \postcode{606-8502}, \state{Kyoto}, \country{Japan}}}

\abstract{Aiming at the relation between QCD and the quark model, 
we consider projections of gauge configurations generated in quenched lattice QCD simulations in the Coulomb gauge on 
a 16$^{\rm 3}$ $\rm \times$ 32, $\rm \beta$ = 6.0 lattice.
First, we focus on a fact that the static quark-antiquark potential is independent of spatial gauge fields.
We explicitly confirm this by performing $\vec{A}$ = 0 projection, where spatial gauge fields are all set to zero.
We also apply the $\vec{A}$ = 0 projection to light hadron masses and find that nucleon and delta baryon masses are almost degenerate, suggesting vanishing of the color-magnetic interactions.
After considering the physical meaning of the $\vec{A}$ = 0 projection,
we next propose a generalized projection, where spatial gauge fields are expanded in terms of Faddeev-Popov eigenmodes and only some eigenmodes are left. 
We apply the proposed projection to light hadron and glueball masses and find that the N-$\rm \Delta$ and 0$^{\rm ++}$-2$^{\rm ++}$ mass splittings become evident when projected with more than 33 (0.10 \%) low-lying eigenmodes, suggesting emergence of the color-magnetic interactions. 
We also find that the original hadron masses are approximately reproduced with just 328 (1.00 \%) low-lying eigenmodes.
These findings indicate an important role of low-lying eigenmodes on hadron masses and would 
be useful in clarifying the relation between QCD and the quark model.}

\maketitle

\section{Introduction}
About 50 years ago, quantum chromodynamics (QCD) was established as the fundamental theory of the strong interaction, because of 
analytical explanation of asymptotic freedom in QCD~\cite{Politzer:1973fx,Gross:1973id}.
Lattice QCD was soon proposed as a non-perturbatively regularized QCD formulated on a space-time  lattice~\cite{Wilson:1974sk}, and the lattice QCD Monte Carlo  simulation was first performed in 1980~\cite{Creutz:1980zw}. 
Since then, lattice QCD simulations have enabled us the first-principles calculation of physical quantities in QCD, 
although it is extremely difficult to analytically solve QCD in the low-energy strong coupling region. 
In the past forty years, many experimental quantities in QCD, such as hadron masses, have been successfully reproduced,
and thus further confirmations of QCD as the fundamental theory of the strong interaction have been provided.

On the other hand, up to now, the quark model has been widely used 
for phenomenological description of various hadrons: 
light mesons and baryons, heavy-light quark systems, 
multi-quark exotic hadrons, inter-hadron interactions and so on
~\cite{Gell-Mann:1964ewy,Zweig:1964ruk,Zweig:1964jf,DeRujula:1975qlm,Oka:1980ax,Oka:1981ri,Godfrey:1985xj,Capstick:1986ter,Loring:2001kx,Vijande:2004he,Valcarce:2005em,Ebert:2007nw,Ebert:2009ub,Vijande:2009kj,Liu:2019zoy,Brambilla:2019esw}.
This model was originally proposed as a classification method of hadrons in terms of constituent quarks \cite{Gell-Mann:1964ewy,Zweig:1964ruk,Zweig:1964jf}
before the establishment of QCD. 
Later the quark model was refined as a theoretical model for the quantitative description of hadrons~\cite{DeRujula:1975qlm,ParticleDataGroup:2022pth}.
The standard quark model is based on a non-relativistic framework, 
and its relativized version of the quark model has been formulated 
and has successfully reproduce light hadron spectra for mesons ~\cite{Godfrey:1985xj,Vijande:2004he,Ebert:2009ub} and baryons~\cite{Capstick:1986ter,Loring:2001kx}. 

Here, unlike the current quark in QCD, 
the constitute quark has a large mass of about 0.3 {\rm GeV}, 
due to which the quark model phenomenologically works 
even for light-quark hadrons in the non-relativistic framework.
Gluonic effects are indirectly included by introducing the static quark-antiquark potential and color-magnetic interactions in the inter consistent-quark potential by hand.
Of course, the quark model is phenomenological and has limitations. 
For example, it cannot describe the quasi Nambu-Goldstone bosons 
(like pions), glueballs and some hadrons with exotic quantum numbers~\cite{ParticleDataGroup:2022pth}.

As a great merit of the quark model, 
it has wide applicability in the hadron physics.
In fact, the quark model has succeeded in describing many properties of wide variety of hadrons: 
both ground-state and excited-state 
light hadrons~\cite{ParticleDataGroup:2022pth},  
heavy-light quark systems~\cite{Ebert:2007nw}, 
multi-quark exotic hadrons~\cite{Vijande:2009kj,Liu:2019zoy,Brambilla:2019esw}, 
and inter-hadron  interactions~\cite{Oka:1980ax,Oka:1981ri,Valcarce:2005em}.
Note here that it is highly difficult to investigate 
highly-excited states of hadrons in lattice QCD simulations. 

Another merit of the quark model is 
to describe the state information with the quark wave function, 
and therefore it is relatively easy to understand the physical process of hadron reactions.
In contrast, it is fairly difficult to extract the state information of each hadron from the lattice QCD simulations, 
because it is based on the path-integral formalism where 
all the possible states are integrated out
~\cite{Suganuma:2023mml}.

Thus, even at present, the quark model is a useful tool, complementary to lattice QCD, and also presents an important phenomenological guide for new aspects in hadron physics.

In spite of the great success of the quark model, its connection with QCD is not yet clear, 
and to clarify the relation between these theories is one of the most important problems in hadron physics.
A significant difference between them is the main dynamical degrees of freedom: while QCD is formulated with current quarks and gluons, the quark model is described only with the massive  constituent quarks.
The other significant difference would be the symmetries they possess.
QCD has local color $\mathrm{SU}(3)$ symmetry, whereas the quark model only has global color $\mathrm{SU}(3)$ symmetry reflecting the absence of dynamical gluons.
This suggests that the quark model is a low-energy effective theory of QCD in some gauge.
This possibility is also supported from studies on the constituent quark mass of $m_{\mathrm{Q}} \simeq 0.3 {\rm GeV}$,
which is a key parameter in the quark model.
In fact, in the Landau gauge, the large quark mass is obtained from the Schwinger-Dyson approach~\cite{Miransky:1983vj,Higashijima:1983gx} and lattice QCD simulations~\cite{Skullerud:2000un,Skullerud:2001aw,Bowman:2002bm,Bonnet:2002ih,Zhang:2003faa,Bowman:2004jm} for the quark propagator. 

Then, what gauge corresponds to the quark model?
The above studies used the Landau gauge mainly because it preserves the Lorentz symmetry in QCD.
However, the non-relativistic quark model does not have 
the whole Lorentz symmetry 
but has only the spatial-rotation symmetry, 
so that the Landau gauge would be out of the candidates.
We consider that gauge to be the Coulomb gauge.
There are several reasons for this choice.
First, the Coulomb gauge leaves global color $\mathrm{SU}(3)$ symmetry at each time slice, which might correspond to the global color symmetry in the quark model.
Second, the Coulomb gauge is globally defined by minimizing spatial gauge-field fluctuations,  
and therefore one expects only a small fluctuation 
of the spatial gluon field, which might explain the absence of dynamical gluons in the quark model.
Third, in the Coulomb gauge, the temporal gauge field $A_0$ 
is no more dynamical but appears as a potential for quarks, 
which might give the static potential in the quark model. 
Several researchers have already studied the relation between the quark model and Coulomb gauge QCD.
For example, Szczepaniak and Swanson studied it in the Hamiltonian formalism~\cite{Szczepaniak:2001rg}.

We study the relation between the quark model and Coulomb gauge QCD in the path-integral formalism by using lattice QCD simulations.
The QCD vacuum has been energetically studied by lattice QCD simulations with projections, such as the Abelian projection~\cite{tHooft:1981bkw,Kronfeld:1987vd,Kronfeld:1987ri} and the center projection~\cite{DelDebbio:1996lih,DelDebbio:1998luz}.
In this paper, we propose a new-type projection in the Coulomb gauge and seek the relation between QCD and the quark model.

The organization of this paper is as follows.
In Sec.~\ref{sec:Coulombgauge}, we briefly review 
QCD in the Coulomb gauge in terms of the Faddeev-Popov mode.
In Sec.~\ref{sec:A0projection}, we focus on a fact that the static quark-antiquark potential is independent of spatial gauge field and propose $\vec{A} = 0$ projection for the gauge configuration generated in lattice QCD simulations. 
In Sec.~\ref{sec:eigenmodeprojection}, we propose a generalization of the $\vec{A} = 0$ projection in the Coulomb gauge utilizing the Faddeev-Popov eigenmodes.
In Sec.~\ref{sec:hadron_mass}, we apply the proposed projection to light hadron and glueball masses. 
Section~\ref{sec:summary} is devoted to Summary and Conclusion.

\section{Coulomb gauge QCD} \label{sec:Coulombgauge}
We briefly review Coulomb gauge QCD.
For the gauge field $A_{s,\mu}=A_{s,\mu}^aT^a \in {\rm su}(N_c)$,
the Coulomb gauge is locally defined as
\begin{equation}
\partial_i A_{s,i}^a = 0.
\end{equation}
In the framework of canonical quantization in the Coulomb gauge, 
the spatial gluon fields $A_{i}^a$ are canonical variables and behave as dynamical variables~\cite{Huang:1982ik}, 
whereas the temporal gluon field $A_0^a$ does not have its canonical momentum field, 
and just behaves as a potential field.

The global definition of the Coulomb gauge is to minimize 
\begin{equation}
R_{\mathrm{C}}[A] := \sum_{i,a} \int dt d^3s\, [A_{s,i}^a(t)]^2
\end{equation}
under the gauge transformation, 
which satisfies the local Coulomb gauge condition.
In fact, the gauge fluctuation of the spatial gluon field 
is strongly suppressed in the Coulomb gauge. 

Thus, the temporal gluon just becomes a potential and spatial gluons are forced to be minimized in the Coulomb gauge.
This situation might provide a preferable matching between QCD and the quark model,
since the quark model does not include dynamical gluons but only has quark degrees of freedom and an inter-quark potential.

In the path-integral formalism, the generating functional of the Yang-Mills (YM) theory 
in the Coulomb gauge is expressed by
\begin{equation}
Z = \int DA\, e^{iS[A]} \delta(\partial_i A_i^a) {\rm Det}(D_i\partial_i)
\end{equation}
with the YM action $S[A]$.
Thus, there appears the Faddeev-Popov (FP) operator, 
\begin{equation}
M^{ab} := D_i^{ab}\partial_i =\partial_i^2\delta^{ab}+gf^{abc}A^c_i \partial_i, 
\label{eq:continuumFP}
\end{equation}
as a weight of the gauge orbital in the Coulomb gauge.
This FP operator is one of the key operators in the Coulomb gauge, 
and its inverse gives the propagator of the FP ghost field.

In lattice formalism, space-time is discretized onto lattice with spacing $a$, 
and the gauge degrees of freedom is described with 
the link variable 
\begin{equation}
U_{s,\mu} := e^{iagA_{s,\mu}} \in {\rm SU}(N_c)
\end{equation}
with the gauge coupling $g$.
In lattice QCD, the global definition of the Coulomb gauge is to maximize 
\begin{equation}
R_{\mathrm{C}} [U] := \sum_{t,s,i} {\rm Re} {\rm Tr} U_{s,i}(t) \label{eq:Coulomb_def}
\end{equation}
under the gauge transformation.
This global definition leads to the local gauge condition
\begin{equation}
\partial^B_i {\cal A}_{s,i} = 0, \label{eq:Coulomb_def_local}
\end{equation}
where $\partial^B$ denotes the backward derivative and 
\begin{equation}
{\cal A}_{s,i} := \frac{1}{2iag}\{U_{s,i}-U_{s,i}^\dagger\}|_{\rm traceless} \in {\rm su}(N_c)
\label{eq:gluonfield}
\end{equation}
goes to the original gluon field $A_i(s)$ in the continuum limit.

In lattice QCD in the Coulomb gauge, the generating functional of the gauge sector is written by
\begin{equation}
Z= \int DU\, e^{-S[U]} \delta(\partial_i {\cal A}_i^a) {\rm Det}(M),
\end{equation}
where the FP operator in the Coulomb gauge takes the form~\cite{Iritani:2012bc} of 
\begin{equation}
M_{x,y}^{a,b} = \sum_{i} A_{x,i}^{a,b} \delta_{x,y} - 
B_{x,i}^{a,b} \delta_{x + \hat{i}, y} - C_{x,i}^{a,b} \delta_{x - \hat{i}, y},
\label{eq:FPlattice}
\end{equation}
with
\begin{align}
A_{x,i}^{a,b} &= {\rm Re}{\rm Tr}[\{T^a, T^b\} ( U_{x,i} + U_{x - \hat{i},i})], \\
B_{x,i}^{a,b} &= 2 {\rm Re}{\rm Tr}[T^b T^a U_{x,i}], \\
C_{x,i}^{a,b} &= 2 {\rm Re}{\rm Tr}[T^a T^b U_{x - \hat{i},i}],
\end{align}
at each time slice, 
which leads to Eq.~(\ref{eq:continuumFP}) in the continuum limit.
Note that $M$ is a real symmetric matrix, i.e.,  
$M^{a,b}_{x,y}=M^{b,a}_{y,x} \in {\bf R}$, 
and hence its eigenvalues $\lambda_n$ are real.

In the Coulomb gauge, this FP operator is a key quantity to control the gauge-orbital weight, and has been studied 
in the context of 
the Gribov horizon~\cite{Gribov:1977wm, Zwanziger:1995cv}, 
the instantaneous color-Coulomb potential~\cite{Zwanziger:1998ez,Szczepaniak:2001rg,Zwanziger:2002sh},
and a confinement scenario, such as the gluon-chain picture~\cite{Greensite:2001nx,tHooft:2002pmx}.

\section{Lattice QCD simulation and $\vec{A} = 0$ projection} \label{sec:A0projection}
We perform SU(3) lattice QCD simulations at the quenched level with the standard plaquette gauge action~\cite{Rothe:1992nt} at $\beta = 6.0$ on a $L_s^3 \times L_t = 16^3 \times 32$ lattice.
We impose periodic boundary conditions for link variables in all directions.
The use of quenched approximation on the relatively small lattice is due to heavy numerical calculations performed for each gauge configuration in Sec.~\ref{sec:eigenmodeprojection}.
We generate 500 gauge configurations, which are picked up every
$500$ sweeps after a thermalization of $5000$ sweeps.
For these gauge configurations, we perform Coulomb gauge fixing by maximizing Eq.~(\ref{eq:Coulomb_def}) under the SU(3) gauge transformation.
We iteratively maximize $R_{\mathrm{C}}[U]$ by successive three SU(2) subgroup gauge transformations.
We stop the iteration so that spatial average of Eq.~(\ref{eq:Coulomb_def_local}) becomes smaller than or an order of $10^{-10}$ at each time slice.
Thus we obtain representatives of the QCD vacuum in the Coulomb gauge.
Our task is to relate them to the quark model.

The static quark-antiquark potential is one of the most fundamental constituents in the quark model, 
and it is usually calculated with the Wilson loop in QCD. 
In principle, however, the static potential can be calculated also from correlator of Polyakov loops, 
which are defined only with temporal gauge fields. 
This indicates that the static quark-antiquark potential is independent of spatial gauge fields at the gauge configuration level. 
%
%
Here, we define $\vec{A} = 0$ projection as a simple replacement of the original gauge configuration in the Coulomb gauge $\{A_{s,\mu}^a\}$ with
\begin{equation}
\{A_{s,\mu}^{a, \mathrm{projected}}  :=(A_{s, 0}^{a}, 0)\}
\end{equation}
and apply to the static quark-antiquark potential.

\begin{figure}[htb]
\includegraphics[width=\linewidth]{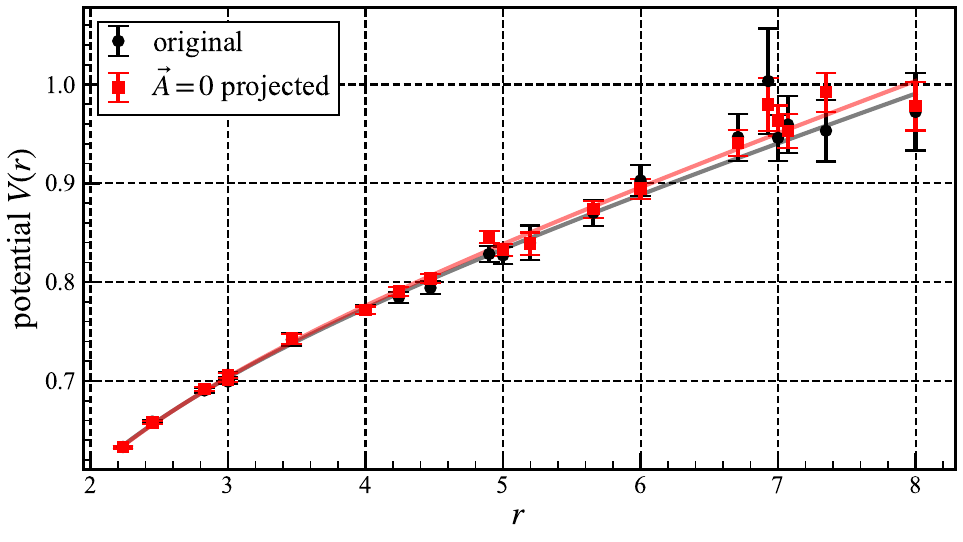}
\caption{\label{fig:qqbar_pot}
The static quark-antiquark potentials evaluated from the original configurations and $\vec{A} = 0$ projected ones in the Coulomb gauge.
The black and red curves are the fit results with the Cornell potential $V(r) = -A / r + \sigma r + C$.
}
\end{figure}

Figure~\ref{fig:qqbar_pot} shows the static quark-antiquark potentials calculated from Wilson loops, 
evaluated from the original gauge configurations and $\vec{A} = 0$ projected ones in the Coulomb gauge. 
Fit results with the Cornell potential are listed in Table.~\ref{table:static_qqbar}.
The static quark-antiquark potential is found to be 
well reproduced even after the $\vec{A} = 0$ projection 
in the Coulomb gauge.
This result is practically nontrivial in lattice QCD, 
since the static quark-antiquark potential is not reproduced after the $\vec{A} = 0$ projection in the numerical lattice QCD calculation 
in the Landau gauge~\cite{Iritani:2011hve}.

\begin{table}[htb]
\caption{
Fit results with the Cornell potential $V(r) = -A / r + \sigma r +C$ for the static quark-antiquark potentials evaluated from the original gauge configurations and $\vec{A} = 0$ projected ones in the Coulomb gauge.
}
\begin{tabular}{cccc}
\hline 
gauge configuration & $A$ & $\sigma$ & $C$ \\
\hline
original                & 0.302(36) & 0.0449(35) & 0.669(23) \\ 
$\vec{A} = 0$ projected & 0.293(28) & 0.0477(25) & 0.658(18) \\
\hline 
\end{tabular}
\label{table:static_qqbar}
\end{table}
\begin{figure*}[htb]
\includegraphics[width=\linewidth]{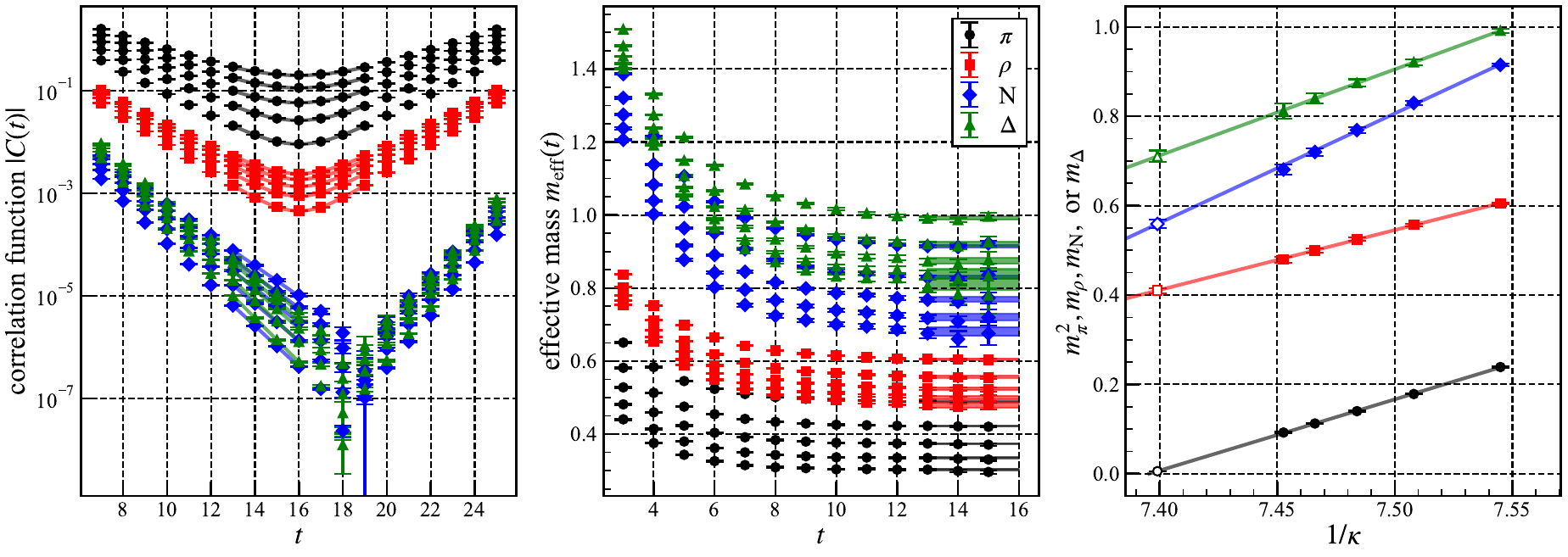}
\caption{\label{fig:hadron_mass_excluding_small_0}
Hadron mass measurement from the original gauge configuration. 
The hopping parameters are taken to be $\kappa = 0.13254, 0.13319, 0.13362, 0.13394, 0.13418$. 
}

\includegraphics[width=\linewidth]{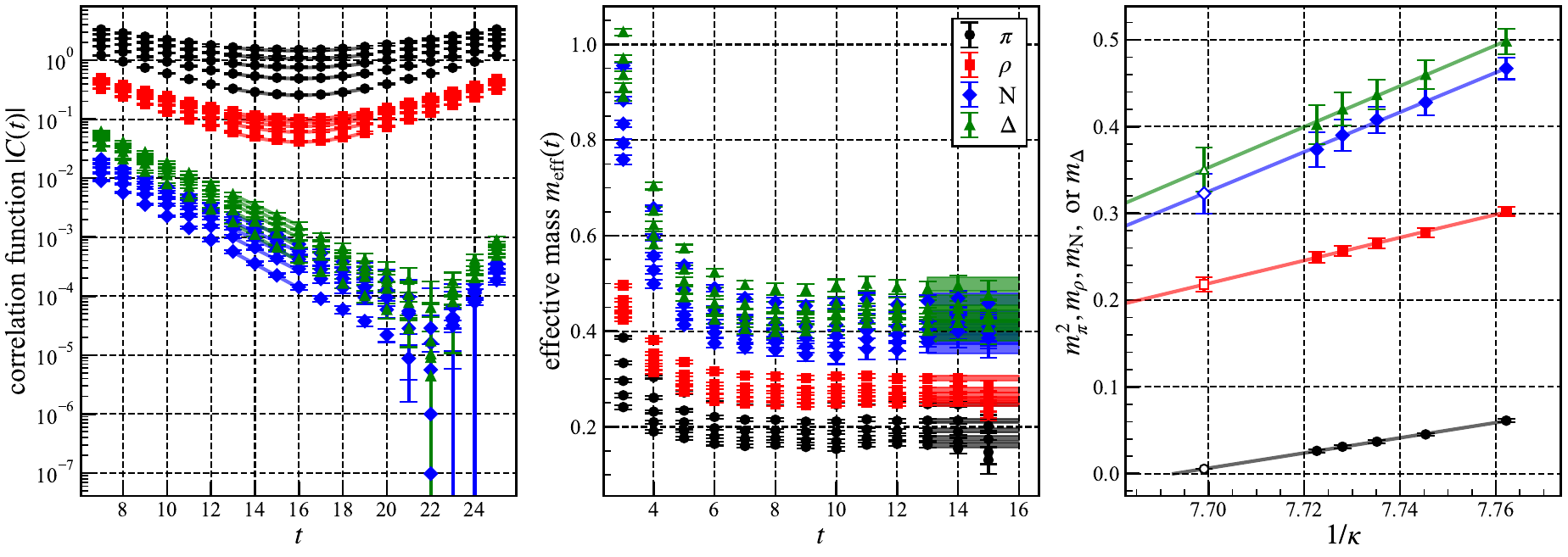}
\caption{\label{fig:hadron_mass_only_small_0}
Hadron mass measurement under the $\vec{A} = 0$ projection
in the Coulomb gauge. 
The hopping parameters are taken to be $\kappa = 0.12883, 0.12911, 0.12928, 0.12940, 0.12949$.
}
\end{figure*}
They agree very well in the whole distance region as expected.

In the quark model, the static quark-antiquark potential is introduced by hand.
In lattice QCD simulations in the Coulomb gauge, we can keep the original static quark-antiquark potential by just leaving the temporal gauge fields unchanged.
Then, the independence of spatial gauge fields can be used to investigate how other fundamental constituents in the quark model, such as the color-magnetic interactions, are encoded in QCD.

Based on the idea, we next apply the $\vec{A} = 0$ projection to hadron masses.
Since hadron masses are not purely static quantities, their values under the $\vec{A} = 0$ projection are quite non-trivial.
We use the clover fermion ($\mathcal{O}(a)$ improved Wilson fermion) with a non-perturbatively determined clover coefficient $c_{\mathrm{SW}} = 1.769$ in the quenched approximation~\cite{Luscher:1996ug}.
For simplicity, we assume that up and down quark masses are degenerate and only consider the lightest hadrons in pseudo scalar, vector, octet, and decuplet sectors, i.e., pion, rho meson, nucleon, and delta baryon.
We extract pion mass $m_\pi$ and rho meson mass $m_\rho$ from correlators defined as
\begin{align}
C(t) &= \sum_{\vec{x}} M^{\mathrm{ud}}(\vec{x}, t) M^{\mathrm{ud}}(\vec{0}, 0)^\dag, \\ 
M^{\mathrm{ud}} &= 
\begin{cases}
\overline{u} \gamma_5 d & \text{for pion}, \\
\overline{u} \gamma_i d & \text{for rho meson}.
\end{cases}
\end{align}
For nucleon mass $m_{\mathrm{N}}$ and delta baryon mass $m_{\Delta}$, we use correlators
\begin{align}
C(t) &= \Gamma_{\alpha \beta} \sum_{\vec{x}} B^{\mathrm{udu}}_{\alpha}(\vec{x},t) \overline{B}_{\beta}^{\mathrm{udu}}(\vec{0},0), \\
B^{\mathrm{udu}} &= 
\begin{cases}
\epsilon^{abc} ((u^{\mathrm{T}})^a C \gamma_5 d^b) u^c \\ 
\qquad \qquad \text{for nucleon}, \\
\epsilon^{abc} 
\{2((u^{\mathrm{T}})^a C \gamma_i d^b) u^c
+ ((u^{\mathrm{T}})^a C \gamma_i u^b) d^c\}
 \\
\qquad \qquad \text{for delta baryon}.
\end{cases}
\end{align}
Here $C = \gamma_0 \gamma_2$ is the charge conjugation matrix, and we set $\Gamma = (1 \pm \gamma_0)$ to extract components proportional to these parity projection operators.
We use exponentially smeared source $\exp(-0.5 |\vec{x}|^2)$ to enhance ground state saturation.
In order to reduce statistical errors, we repeat the measurement for two distant hadron-source locations and take the average for each gauge configuration.
In the analysis, we exclude exceptional gauge configurations, where the pion correlator has ten times larger value than the statistical average.

Figure~\ref{fig:hadron_mass_excluding_small_0} shows light hadron mass measurement from the original gauge configuration.
For each hoppoing parameter $\kappa$ and sector, hadron mass is obtained from single cosh/exponential fit to the correlation function.
The fit range is determined from the effective mass plot.
From linear chiral extrapolations of $m_{\pi}^2$ and $m_{\rho}$, 
the physical hopping parameter and the lattice spacing are determined so that $m_{\pi} = 0.14 {\rm GeV}$ and $m_{\rho} = 0.77 {\rm GeV}$.
The lattice spacing is found to be $a = 0.1051(15) {\rm fm}$.
Nucleon and delta baryon masses are found to be $m_{\mathrm{N}} = 1.050(24) {\rm GeV}$ and $m_{\Delta} = 1.334(31) {\rm GeV}$, 
which are consistent with a quenched lattice QCD simulation with a similar lattice setup~\cite{JLQCD:2002zto} (see Table~XXI).

On the other hand, Fig.~\ref{fig:hadron_mass_only_small_0} shows light hadron mass measurement under the $\vec{A} = 0$ projection. 
Here we assume that the lattice spacing is the same as the one obtained from the original gauge configuration because the static quark-antiquark potential is unchanged.
The physical hopping parameter is determined so that $m_{\pi} = 0.14 {\rm GeV}$.
Rho meson, nucleon, and delta baryon masses are found to be $m_{\rho} = 0.409(17) {\rm GeV}, m_{\mathrm{N}} = 0.607(44){\rm GeV}$, and $m_{\Delta} = 0.658(49) {\rm GeV}$, respectively under the $\vec{A} = 0$ projection.
The overall decrease of hadron masses suggests a constituent quark with smaller mass of $m_{\mathrm{Q}} \simeq 0.2 {\rm GeV}$.
Interestingly, we find that nucleon and delta baryon masses are almost degenerate under the $\vec{A} = 0$ projection.
Since the $\mathrm{N}-\Delta$ mass splitting is generated by the color-magnetic interactions in the quark model,
it suggests vanishing of such interactions under the $\vec{A} = 0$ projection.

Here we consider the physical meaning of the $\vec{A} = 0$ projection.
Under the $\vec{A} = 0$ projection, color-magnetic fields 
$H_i^a=\epsilon_{ijk}(\partial_j A_k^a
-\partial_k A_j^a
-g f^{abc}A_j^b A_k^c)$
are inevitably zero, while color-electric fields can remain.
Hence, it is natural to consider that color-magnetic interactions vanish under the $\vec{A} = 0$ projection.
Because only the color-magnetic interactions are spin-dependent for static systems in QCD, 
mass splitting between different spin states are universally expected to vanish or to decrease under the $\vec{A} = 0$ projection.

In order to check the conjecture,
we also want to obtain $0^{++}$ and $2^{++}$ glueball masses under the $\vec{A} = 0$ projection.
However, it seems to be difficult.
These masses can be extracted from correlators
\begin{align}
C(t) &= \phi(t) \phi(0), \label{eq:glueball_corr1} \\ 
\phi(t) &= 
\begin{cases}
\sum_{\vec{x}} {\rm Re} {\rm tr}(P_{12} + P_{23} + P_{31})(\vec{x}, t) & \text{for $0^{++}$}, \\
\sum_{\vec{x}} {\rm Re} {\rm tr}(P_{12} - P_{13})(\vec{x}, t) & \text{for $2^{++}$},
\end{cases}
\label{eq:glueball_corr2}
\end{align}
where $P_{ij}$ is the plaquette in $i-j$ plane and the vacuum contribution must be subtracted from the correlator for $0^{++}$ glueball mass~\cite{Ishikawa:1982tb}.
Since these correlators are exactly zero under the $\vec{A} = 0$ projection, we cannot obtain temporal dependencies of them.
A possible way to evaluate $0^{++}$ and $2^{++}$ glueball masses under the $\vec{A} = 0$ projection
would be to take a limit of a projection which smoothly connects the $\vec{A} = 0$ projected configuration with the original one.
Such projection would also be useful to investigate how physical quantities under the $\vec{A} = 0$ projection and the original ones are related.
For these purposes, we develop a generalization of the $\vec{A} = 0$ projection in the next section.

\section{A generalization of the $\vec{A} = 0$ projection in the Coulomb gauge} \label{sec:eigenmodeprojection}
We utilize the Faddeev-Popov (FP) operator $M$ to develop a generalization of the $\vec{A} = 0$ projection, 
since $M$ is considered to be an important ingredient in Coulomb gauge QCD,
as briefly reviewed in Sec.~\ref{sec:Coulombgauge}.
Roughly, the FP operator defined by Eq.~(\ref{eq:continuumFP}) 
resembles the Laplacian in the Coulomb gauge, 
where spatial gluon amplitude is fairly suppressed, 
and low-lying FP modes correspond to low-energy components.
In lattice QCD, the FP operator $M$ 
is expressed as a $8L_s^3 \times 8L_s^3$ matrix $M_{IJ}=M_{x,y}^{a,b}$ in Eq.~(\ref{eq:FPlattice}), 
with the index $I=(x,a)$ and $J=(y,b)$.

We consider expansion of spatial gauge fields in terms of the FP eigenmodes.
Since the FP operator is a real symmetric matrix, its eigenmodes 
\begin{equation}
M \psi_n = \lambda_n \psi_n,\, n = 1, 2, \dots, 8L_s^3
\end{equation}
can be taken real and normalized eigenvectors $\psi_n(t)$ at a time $t$ form a complete set 
\begin{equation}
\sum_{n}\psi_{n,s}^a(t) \psi_{n,s^{\prime}}^{a^{\prime}}(t) = \delta_{s,s^\prime} \delta^{a, a^{\prime}}.
\end{equation} 
With these eigenvectors, spatial gauge fields at a time $t$ can be uniquely expanded as
\begin{align}
A_{s,i}^a(t) &= \sum_{n = 1}^{8 L_s^3} c_{n,i}(t) \psi_{n,s}^a(t), \\
c_{n,i}(t) & := \sum_{s,a} A_{s,i}^a(t) \psi_{n,s}^a(t).
\end{align}
Among the $8L_s^3$ eigenmodes, we consider a restriction of these modes.
We here define \textit{eigenmode projection} as a replacement of the original gauge configuration in the Coulomb gauge $\{A_{s, \mu}^a\}$ with
\begin{equation}
\{A_{s,\mu}^{a,\mathrm{projected}} := (A_{s, 0}^a, \sum_{n \in S} c_{n,i} \psi_{n,s}^a)\},
\end{equation}
where $S$ denotes a subset of the whole set $\{1,\dots, 8L_s^3\}$.
The eigenmode projection with no eigenmode ($S = \emptyset$) is the $\vec{A} = 0$ projection.
On the other hand, if we set $S = \{1, \dots, 8L_s^3\}$, 
the eigenmode projection does not change the configuration at all.
We can connect the $\vec{A} = 0$ projected configuration with the original one by smoothly changing the subset $S$. 


In lattice QCD, the gluon field 
$A_{s,\mu}^{a,\mathrm{projected}} \in {\rm SU}(3)$ 
is converted into the corresponding link-variable 
$U_{s,\mu}^{a,\mathrm{projected}}:=\exp{(iag A_{s,\mu}^{a,\mathrm{projected}})} \in {\rm SU}(3)$ 
on each link, where a standard diagonalization technique of a matrix is used.
Then, the link-variable $U_{s,\mu}^{a,\mathrm{projected}}$ 
is used for the following calculations 
on the lattice. 



In this paper, we only consider eigenmode projections with $N_{\mathrm{low}}$ low-lying or $N_{\mathrm{high}}$ high-lying eigenmodes.
Specifically, we set
\begin{align}
S_{\mathrm{low}} &= \{1, \dots, N_{\mathrm{low}}\}, \\
S_{\mathrm{high}} &= \{8L_s^3 - N_{\mathrm{high}} + 1, \dots, 8L_s^3\}
\end{align}
for eigenmodes which are ordered as
\begin{equation}
\lambda_1 \le \lambda_2 \le \dots \le \lambda_{8L_s^3}.
\end{equation}
For each gauge configuration and time $t$, we calculate 1638 low-lying and high-lying eigenmodes for the FP operator by using the Krylov-Schur solver implemented in SLEPc~\cite{Hernandez:2005:SSF}.
In fact, we solve the large-scale eigenvalue problem $500 \times 32 \times 2$ times.
Since the total number of eigenmodes is $8L_s^3 = 32768$, we can cover $0-5.00\, \%$ and also $95.00-100\, \%$ of the whole range.
Note that, for example, the gauge configuration projected with 31130 (95.00 \%) low-lying eigenmodes can be obtained by subtracting that projected with 1638 (5.00 \%) high-lying eigenmodes from the original one.
\begin{figure}[htb]
\includegraphics[width=\linewidth]{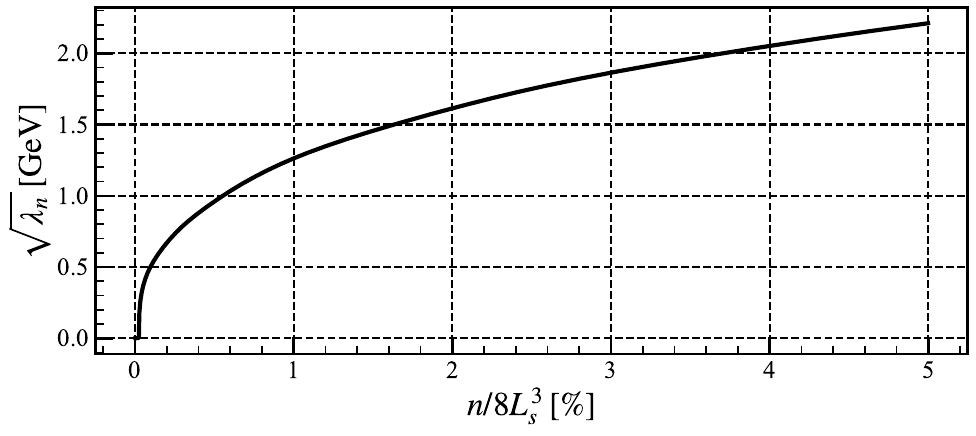}
\caption{\label{fig:eigenvalue}
Root of the $n$-th smallest eigenvalue $\sqrt{\lambda_n}$ in GeV unit plotted against $n / 8L_s^3$. 
The curve is made from lines connecting adjacent data points.
The lattice spacing $a = 0.1051 {\rm fm}$ is used to set the scale.
}
\end{figure}
Figure~\ref{fig:eigenvalue} shows the $n$-th smallest eigenvalue $\lambda_n$ as a function of $n / 8L_s^3$.
It is found that the 1638-th smallest eigenvalue corresponds to the low-lying eigenmode of about $(2 {\rm GeV})^2$.

Instead of the expansion in terms of the FP eigenmodes,
one might prefer the Fourier expansion~\cite{Yamamoto:2008am,Yamamoto:2008ze}. 
However, the Laplacian has so many degenerate eigenmodes that it is not easy to change the subset $S$ smoothly.
In the current expansion,
such degeneracy of eigenmodes practically does not exist except for the trivial eight zero-modes, 
and one can smoothly connect the $\vec{A} = 0$ projected gauge configuration with the original one.

\section{hadron and glueball masses under the eigenmode projection} \label{sec:hadron_mass}
We calculate light hadron masses under the eigenmode projection in the same way as was done in Sec.~\ref{sec:A0projection}.
We summarize hopping parameters and corresponding $m_{\pi} / m_{\rho}$ in the Appendix.
Figure~\ref{fig:hadron_mass} shows light hadron masses under the eigenmodes projection with low-lying or high-lying eigenmodes.
Some of these results are listed in Table~\ref{table:hadron_mass}.
\begin{table}[htb]
\caption{
Hadron masses under the eigenmode projection with low-lying FP eigenmodes.
}
\begin{tabular}{cccc}
\hline
num. of low-lying FP modes & $m_{\rho}$ [GeV] & $m_{\mathrm{N}}$ [GeV] & $m_{\Delta}$ [GeV] \\
\hline
0 (0 \%)          & 0.409(17)    & 0.607(44) & 0.658(49) \\ 
8 (0.02 \%)       & 0.569(23)    & 0.788(86) & 0.887(68) \\
33 (0.10 \%)      & 0.675(25)    & 0.984(68) & 1.122(77) \\
164 (0.50 \%)     & 0.772(20)    & 1.045(35) & 1.306(50) \\
328 (1.00 \%)     & 0.785(21)    & 1.051(36) & 1.372(54) \\
$8L_s^3$ (100 \%) & 0.77         & 1.050(24) & 1.334(31) \\
\hline 
\end{tabular}
\label{table:hadron_mass}
\end{table}
We also show hadron mass measurements under the eigenmode projection with 1638 (5.00 \%) low-lying and 31130 (95.00 \%) high-lying eigenmodes in Figs.~\ref{fig:hadron_mass_only_small_1638} and \ref{fig:hadron_mass_excluding_small_1638}.
\begin{figure*}[htb]
\includegraphics[width=\linewidth]{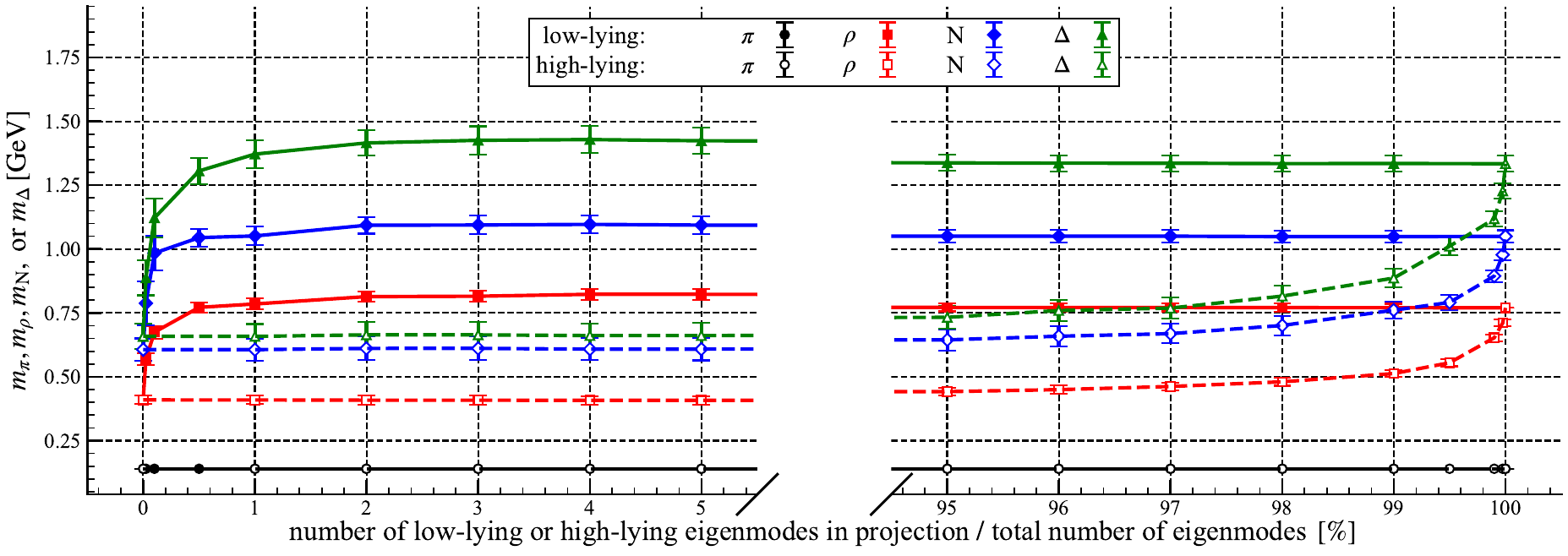}
\caption{\label{fig:hadron_mass}
Hadron masses under the eigenmode projection in GeV unit plotted against the number of low-lying or high-lying FP eigenmodes in the projection. 
Pion mass $m_\pi = 0.14 {\rm GeV}$ is used to set the physical hopping parameter for each measurement. 
The lattice spacing $a = 0.1051(15) {\rm fm}$ is used to set the scale.
The solid and dashed lines are connecting adjacent data points for visibility. 
}
\includegraphics[width=\linewidth]{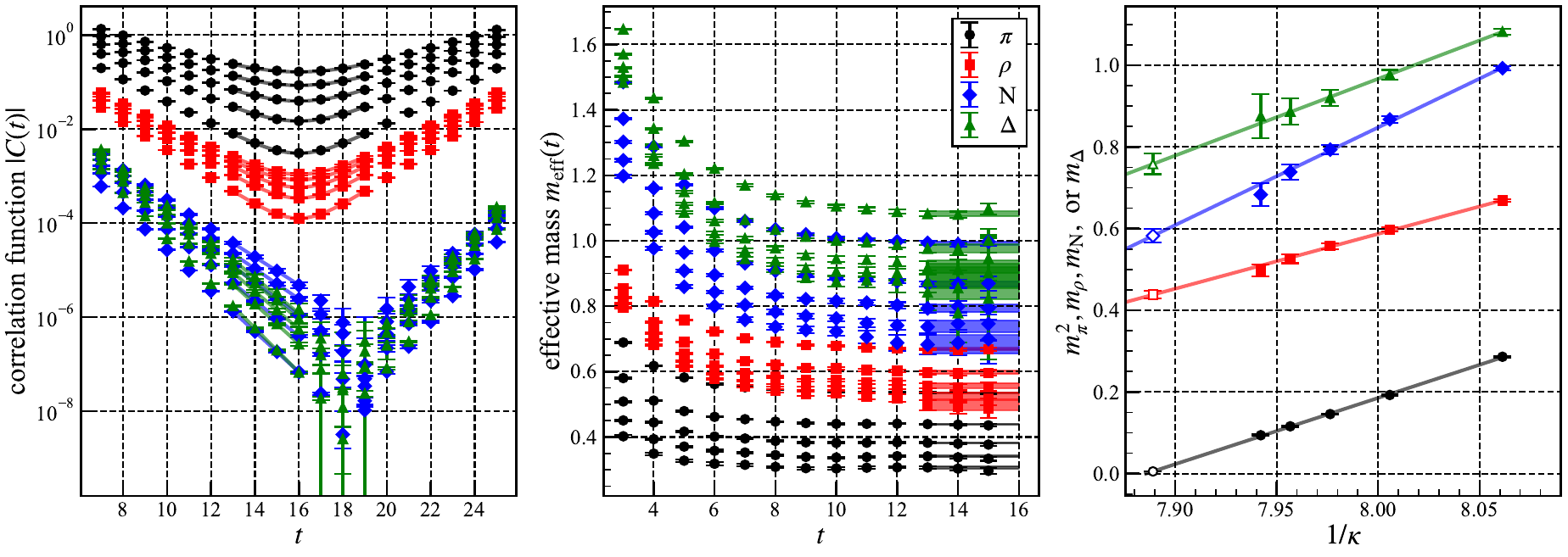}
\caption{\label{fig:hadron_mass_only_small_1638}
Hadron mass measurement under the eigenmode projection with 1638 (5.00 \%) low-lying FP eigenmodes.
The hopping parameters are taken to be $\kappa = 0.12405, 0.12491, 0.12537, 0.12568, 0.12591$.
}

\includegraphics[width=\linewidth]{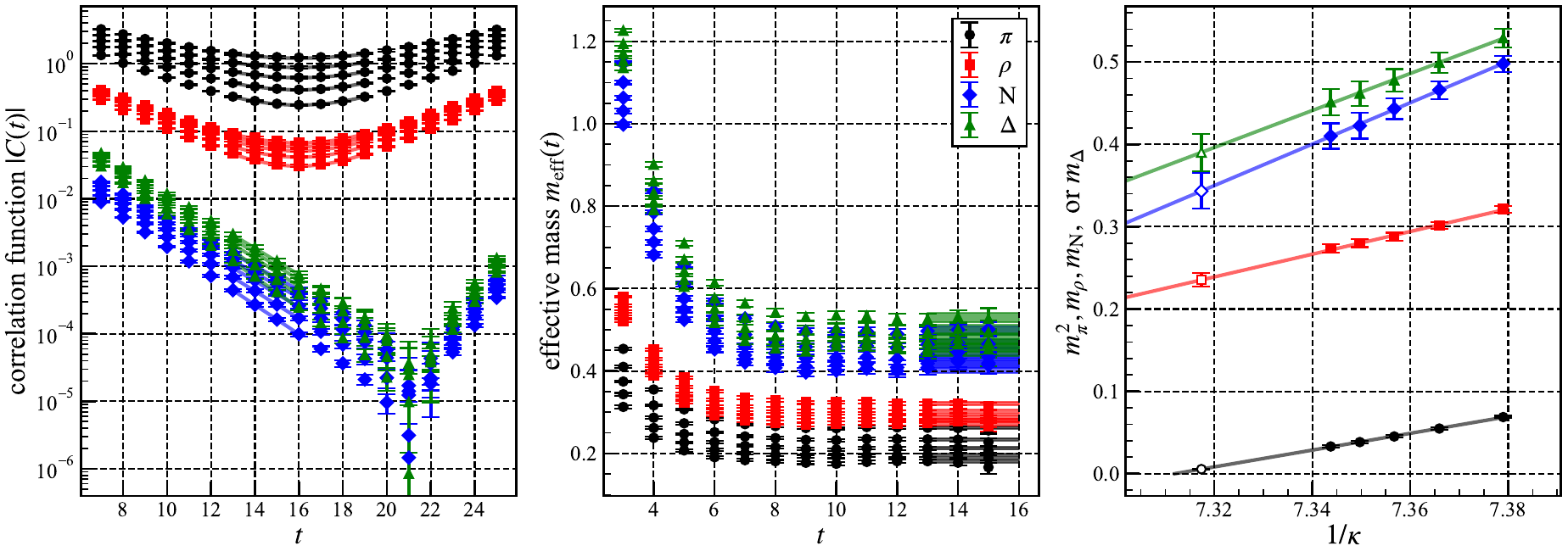}
\caption{\label{fig:hadron_mass_excluding_small_1638}
Hadron mass measurement under the eigenmode projection with 31130 (95.00 \%) high-lying FP eigenmodes.
The hopping parameters are taken to be $\kappa = 0.13552, 0.13576, 0.13593, 0.13606, 0.13617$.
}
\end{figure*}

We find monotonically increasing behaviors in Fig.~\ref{fig:hadron_mass} although slight discrepancies exist,
which probably appeared due to uncertainties in the chiral extrapolations.
However, the speed of convergence is completely different: the original light hadron masses are approximately reproduced with only one percent of low-lying eigenmodes, whereas they can hardly be reproduced with high-lying eigenmodes.
Interestingly, Figs.~\ref{fig:hadron_mass_only_small_1638} and \ref{fig:hadron_mass_excluding_small_1638} are respectively very similar 
with Figs.~\ref{fig:hadron_mass_excluding_small_0} and \ref{fig:hadron_mass_only_small_0} in Sec.~\ref{sec:A0projection}, 
except for the values of hopping parameter $\kappa$.
These results indicate an important role of low-lying eigenmodes on light hadron masses.
From Fig.~\ref{fig:hadron_mass} and Table~\ref{table:hadron_mass}, 
we also find that the $\mathrm{N}-\Delta$ mass splitting becomes evident when projected with more than 33 (0.10 \%) low-lying eigenmodes.

\begin{figure*}[htb]
\includegraphics[width=\linewidth]{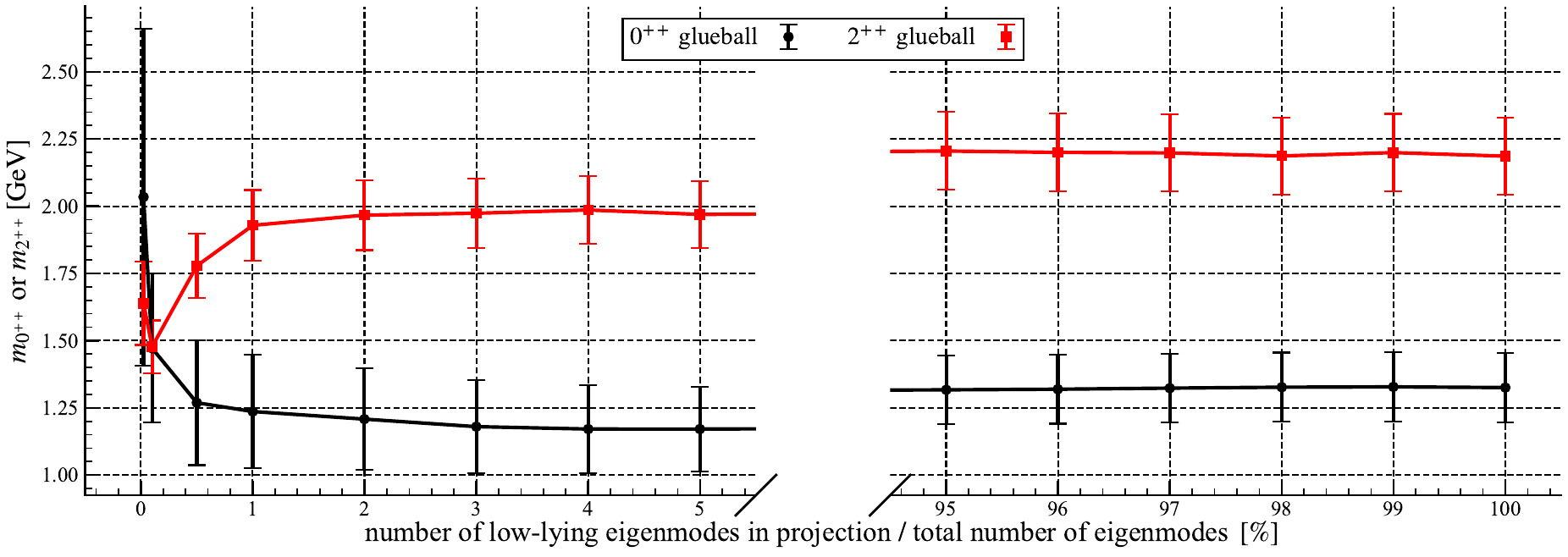}
\caption{\label{fig:glueball_mass}
Glueball masses under the eigenmode projection in GeV unit plotted against the number of low-lying FP eigenmodes in the projection.
The lattice spacing $a = 0.1051(15) {\rm fm}$ is used to set the scale.
$0^{++}$ and $2^{++}$ glueball masses are extracted from single exponential fits to the correlation functions with the fit ranges of $[2, 5]$ and $[1, 5]$, respectively.
}
\end{figure*}

\begin{figure*}[htb]
\includegraphics[width=\linewidth]{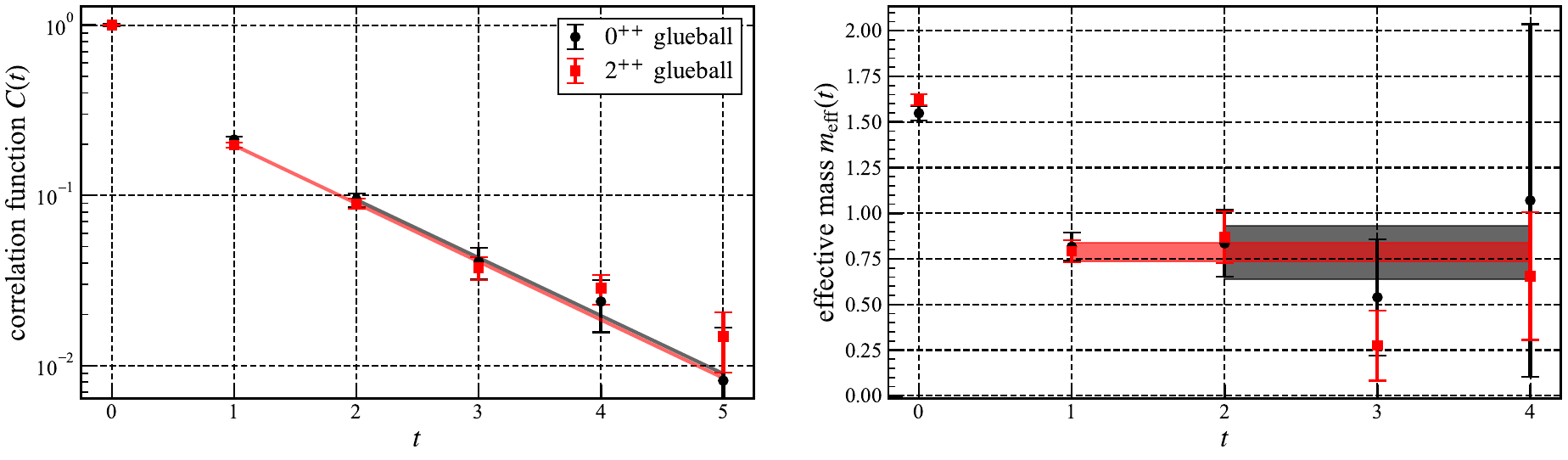}
\caption{\label{fig:glueball_mass_only_small_33}
Glueball mass measurement under the eigenmode projection with 33 (0.10 \%) low-lying FP eigenmodes.
Correlation functions are normalized such that $C(0) = 1$.
}
\end{figure*}

We next apply the eigenmode projection to $0^{++}$ and $2^{++}$ glueball masses.
It is known that a large number of gauge configurations is needed for the conclusive glueball mass calculations in lattice QCD simulations.
Although we enhance the signals by performing the three times blocking of Teper~\cite{Teper:1987wt}, 500 gauge configurations are obviously not enough.
Nevertheless, we succeed in calculating approximate $0^{++}$ and $2^{++}$ glueball masses under the eigenmode projection with low-lying eigenmodes,
from single exponential fits to the correlators~(\ref{eq:glueball_corr1}, \ref{eq:glueball_corr2}) with the fit ranges of $[2, 5]$ and $[1, 5]$, respectively.
Those results are summarized in Fig.~\ref{fig:glueball_mass}.
We also show glueball mass mesurement under the eigenmode projection with 33 (0.10 \%) low-lying eigenmodes in Fig.~\ref{fig:glueball_mass_only_small_33}.
In the case of projection with high-lying eigenmodes, 
the correlation functions are too noisy that we can not extract the masses at all.

From Fig.~\ref{fig:glueball_mass}, it seems that $0^{++}$ and $2^{++}$ glueball masses are approximately degenerate near the $\vec{A} = 0$ projection.
This can be also seen from Fig.~\ref{fig:glueball_mass_only_small_33}, where correlation functions are very alike.
They approximately converge to the original values with just one percent of low-lying eigenmodes.
We also find that the $0^{++}-2^{++}$ glueball mass splitting becomes evident when projected with more than 33 (0.10 \%) low-lying eigenmodes.
Thus, we obtain similar results also in the case of glueball masses.

\section{Summary and Conclusion} \label{sec:summary}
In this paper, aiming at the relation between QCD and the quark model, 
we have performed projections of gauge configurations generated in quenched lattice QCD simulations in the Coulomb gauge on 
a $16^3 \times 32, \beta = 6.0$ lattice.
First, we have focused on a fact that the static quark-antiquark potential is independent of spatial gauge fields and explicitly confirmed it by the $\vec{A} = 0$ projection.
We have also applied the $\vec{A} = 0$ projection to light hadron masses and found that nucleon and delta baryon masses are almost degenerate.
We have given an interpretation of this by vanishing of the color-magnetic interactions under the $\vec{A} = 0$ projection.

Next, we have proposed the eigenmode projection as a generalization of the $\vec{A} = 0$ projection and applied to light hadron and glueball masses.
We have obtained similar results both for light hadron and glueball masses: mass splitting between different spin states approximately vanishes near the $\vec{A} = 0$ projection and the original masses are reproduced with just 328 (1.00 \%) low-lying eigenmodes.
We have also found that the $N-\Delta$ and $0^{++}-2^{++}$ mass splittings become evident when projected with more than 33 (0.10 \%) low-lying eigenmodes, suggesting emergence of the color-magnetic interactions. 

The eigenmode projection has introduced a new perspective in the analysis of low-energy phenomena in QCD, namely the relevant eigenmodes.
From Fig.~\ref{fig:eigenvalue}, we can see that the 33-th and 328-th smallest eigenvalues correspond to low-lying eigenmodes of about $(0.5 {\rm GeV})^2$ and $(1.3 {\rm GeV})^2$, respectively.
These energies might correspond to the energy scales of color-magnetic interactions and constituent quark mass generation, respectively.
Our study suggests a possibility that the quark model can be derived from Coulomb gauge QCD by a reduction of high-lying spatial gluons above $1.3{\rm GeV}$.
Also, a QCD-based constituent gluon model 
might be able to be formulated from Coulomb gauge QCD with the reduction of high-lying spatial gluons, and its combination with the quark model might describe wide category of hadrons including glueballs and hybrids.

In this way, we find some remnants of the quark model in the QCD vacuum generated from lattice QCD simulations in the Coulomb gauge. 
We expect that the Coulomb gauge is one of the suitable gauge to connect the quark model from QCD.  

Finally, we mention remaining questions and related future works.
Still now, the quark model has a difficult puzzle as to 
the physical reason why the non-relativistic potential model  phenomenologically works even in the light quark sector. 
Although some relativized version of the quark model has improved the kinetic term of quarks \cite{Godfrey:1985xj,Capstick:1986ter, Loring:2001kx,Ebert:2007nw,Ebert:2009ub}, it is still based on the potential picture. 
It is difficult to imagine that such a naive quark potential model works for light hadrons because non-relativistic effect should appear for light quarks.
A physical reason of the phenomenological success of the quark potential model might be a large constituent quark mass of about 0.3 {\rm GeV} in the light-quark sector, reflecting spontaneous chiral symmetry breaking. 

Then, as a future work, it is necessary to check the effects of dynamical quarks using the full lattice QCD simulations.
In particular, it would be interesting to evaluate the effect of quark motion with changing the current quark mass, 
since the quark potential picture works for heavy quarks 
and it might show some discrepancy for light quarks.

In this paper, we have shown that 
only low-lying modes of the Faddeev-Popov operator $D_i \partial_i$ 
give the relevant contribution on hadron masses. 
In other words, all the high-lying Faddeev-Popov modes 
do not seem to contribute the hadron masses. 
Then, it is an interesting challenge to construct 
some effective field theory only with the low-lying Faddeev-Popov modes starting from QCD, since it might be an intermediate theory connecting between QCD and the quark model.
However, such an attempt would be highly difficult in the analytical manner. 
%

In this study, we have focused the Faddeev-Popov operator 
in the Coulomb gauge. 
It is also desirable to study the eigenmode projection with different operators, such as the Laplacian, to check whether the Faddeev-Popov operator is special or not in the current approach.
As an another future study, it would be interesting to study eigenmode projection in the Landau gauge.
In the Landau gauge, the Lorentz symmetry is preserved and eigenmode projection can naturally connect $A_{\mu} = 0$ projected configuration with the original one.
There, one can analyze how the original static quark-antiquark potential is reproduced.
Such study would contribute our understanding on the mechanism of quark confinement.

\bmhead{Acknowledgments}
H.O. was supported by a Grant-in-Aid 
for JSPS Fellows (Grant No.21J20089).
H.S. was supported by a Grants-in-Aid for
Scientific Research [19K03869] from Japan Society for the Promotion of Science. 
This work was in part based on Bridge++ code~\cite{Ueda:2014rya}.
We have used SLEPc~\cite{Hernandez:2005:SSF} to solve eigenvalue problems for the Faddeev-Popov operator.
The numerical simulations have been carried out on Yukawa-21 at Yukawa Institute for Theoretical Physics, Kyoto University.

\begin{appendices}
\section{hopping parameters} \label{sec:appendix}
The hopping parameters used in hadron mass measurements under the eigenmode projection remaining low/high-lying FP modes are listed in Table~\ref{table:low-hopping} and 
Table~\ref{table:high-hopping}, respectively.
We roughly tuned them to cover the range of $m_{\pi} / m_{\rho} = 0.80 - 0.60$.
The resulting $m_{\pi} / m_{\rho}$ are also listed.
\begin{table*}[htb]
\caption{
Hopping parameters and corresponding $m_{\pi} / m_{\rho}$ in the light hadron mass measurement under the eigenmode projection remaining low-lying FP modes.
}
\begin{tabular}{ccccccc}
\hline
\# of low modes \\
\hline
0 (0 \%) & $\kappa$ &
0.12883 & 0.12911 & 0.12928 & 0.12940 & 0.12949 \\
& $m_{\pi} / m_{\rho}$ &
0.819(18) & 0.766(21) & 0.724(23) & 0.686(25) & 0.650(27) \\
8 (0.02 \%) & $\kappa$ &
0.12853 & 0.12879 & 0.12899 & 0.12915 & 0.12928 \\
& $m_{\pi} / m_{\rho}$ &
0.796(13) & 0.754(15) & 0.712(17) & 0.669(19) & 0.634(22) \\
33 (0.10 \%) & $\kappa$ &
0.12774 & 0.12818 & 0.12845 & 0.12864 & 0.12879 \\
& $m_{\pi} / m_{\rho}$ &
0.788(10) & 0.732(13) & 0.683(15) & 0.639(18) & 0.598(22) \\
164 (0.50 \%) & $\kappa$ &
0.12564 & 0.12647 & 0.12696 & 0.12731 & 0.12757 \\
& $m_{\pi} / m_{\rho}$ &
0.8160(54) & 0.7552(76) & 0.705(10) & 0.659(14) & 0.618(18) \\
328 (1.00 \%) & $\kappa$ &
0.12510 & 0.12592 & 0.12649 & 0.12686 & 0.12712 \\
& $m_{\pi} / m_{\rho}$ &
0.8070(49) & 0.7476(72) & 0.690(11) & 0.647(15) & 0.618(21) \\
655 (2.00 \%) & $\kappa$ &
0.12448 & 0.12535 & 0.12590 & 0.12627 & 0.12654 \\
& $m_{\pi} / m_{\rho}$ &
0.8094(44) & 0.7492(67) & 0.6955(98) & 0.652(14) & 0.618(20) \\
983 (3.00 \%) & $\kappa$ &
0.12443 & 0.12523 & 0.12568 & 0.12599 & 0.12623 \\
& $m_{\pi} / m_{\rho}$ &
0.7950(47) & 0.7334(73) & 0.687(10) & 0.649(14) & 0.618(20) \\
1311 (4.00 \%) & $\kappa$ &
0.12424 & 0.12505 & 0.12550 & 0.12581 & 0.12604 \\
& $m_{\pi} / m_{\rho}$ &
0.7951(46) & 0.7316(72) & 0.684(10) & 0.645(14) & 0.615(20) \\
1638 (5.00 \%) & $\kappa$ &
0.12405 & 0.12491 & 0.12537 & 0.12568 & 0.12591 \\
& $m_{\pi} / m_{\rho}$ &
0.8000(44) & 0.7346(69) & 0.685(10) & 0.648(14) & 0.616(20) \\
31130 (95.00 \%) & $\kappa$ &
0.13214 & 0.13279 & 0.13322 & 0.13353 & 0.13377 \\
& $m_{\pi} / m_{\rho}$ &
0.8065(33) & 0.7572(45) & 0.7121(59) & 0.6706(77) & 0.633(10) \\
31457 (96.00 \%) & $\kappa$ &
0.13221 & 0.13286 & 0.13329 & 0.13361 & 0.13386 \\
& $m_{\pi} / m_{\rho}$ &
0.8075(32) & 0.7587(44) & 0.7139(59) & 0.6714(77) & 0.632(10) \\
31785 (97.00 \%) & $\kappa$ &
0.13229 & 0.13295 & 0.13338 & 0.13370 & 0.13394 \\
& $m_{\pi} / m_{\rho}$ &
0.8077(32) & 0.7581(44) & 0.7132(59) & 0.6705(77) & 0.6323(99) \\
32113 (98.00 \%) & $\kappa$ &
0.13238 & 0.13303 & 0.13346 & 0.13378 & 0.13402 \\
& $m_{\pi} / m_{\rho}$ &
0.8073(32) & 0.7583(44) & 0.7136(59) & 0.6711(77) & 0.6332(99) \\
32440 (99.00 \%) & $\kappa$ &
0.13246 & 0.13311 & 0.13355 & 0.13386 & 0.13411 \\
& $m_{\pi} / m_{\rho}$ &
0.8073(32) & 0.7584(44) & 0.7124(59) & 0.6711(77) & 0.631(10) \\
$8L_s^3$ (100 \%) & $\kappa$ &
0.13254 & 0.13319 & 0.13362 & 0.13394 & 0.13418 \\
& $m_{\pi} / m_{\rho}$ &
0.8071(32) & 0.7582(44) & 0.7134(59) & 0.6710(77) & 0.6332(99) \\
\hline 
\end{tabular}
\label{table:low-hopping}
\end{table*}

\begin{table*}[htb]
\caption{
Hopping parameters and corresponding $m_{\pi} / m_{\rho}$ in the light hadron mass measurement under the eigenmode projection remaining high-lying FP modes.
}
\begin{tabular}{ccccccc}
\hline
\# of high modes \\
\hline
0 (0 \%) & $\kappa$ &
0.12883 & 0.12911 & 0.12928 & 0.12940 & 0.12949 \\
& $m_{\pi} / m_{\rho}$ &
0.819(18) & 0.766(21) & 0.724(23) & 0.686(25) & 0.650(27) \\
328 (1.00 \%) & $\kappa$ &
0.12887 & 0.12916 & 0.12933 & 0.12945 & 0.12954 \\
& $m_{\pi} / m_{\rho}$ &
0.820(18) & 0.766(21) & 0.723(23) & 0.686(25) & 0.650(27) \\
655 (2.00 \%) & $\kappa$ &
0.12894 & 0.12919 & 0.12935 & 0.12947 & 0.12956 \\
& $m_{\pi} / m_{\rho}$ &
0.817(18) & 0.771(20) & 0.732(22) & 0.696(24) & 0.664(26) \\
983 (3.00 \%) & $\kappa$ &
0.12899 & 0.12924 & 0.12940 & 0.12952 & 0.12961 \\
& $m_{\pi} / m_{\rho}$ &
0.817(18) & 0.771(20) & 0.732(22) & 0.697(24) & 0.665(26) \\
1311 (4.00 \%) & $\kappa$ &
0.12901 & 0.12929 & 0.12946 & 0.12959 & 0.12968 \\
& $m_{\pi} / m_{\rho}$ &
0.822(18) & 0.771(20) & 0.730(22) & 0.691(24) & 0.656(26) \\
1638 (5.00 \%) & $\kappa$ &
0.12908 & 0.12935 & 0.12952 & 0.12964 & 0.12973 \\
& $m_{\pi} / m_{\rho}$ &
0.819(18) & 0.769(21) & 0.727(23) & 0.691(24) & 0.657(26) \\
31130 (95.00 \%) & $\kappa$ &
0.13552 & 0.13576 & 0.13593 & 0.13606 & 0.13617 \\
& $m_{\pi} / m_{\rho}$ &
0.816(14) & 0.777(15) & 0.738(17) & 0.700(18) & 0.665(20) \\
31457 (96.00 \%) & $\kappa$ &
0.13550 & 0.13575 & 0.13593 & 0.13607 & 0.13618 \\
& $m_{\pi} / m_{\rho}$ &
0.816(13) & 0.775(15) & 0.734(17) & 0.696(17) & 0.656(20) \\
31785 (97.00 \%) & $\kappa$ &
0.13545 & 0.13571 & 0.13590 & 0.13605 & 0.13616 \\
& $m_{\pi} / m_{\rho}$ &
0.814(12) & 0.772(14) & 0.729(16) & 0.683(17) & 0.644(20) \\
32113 (98.00 \%) & $\kappa$ &
0.13534 & 0.13560 & 0.13579 & 0.13594 & 0.13605 \\
& $m_{\pi} / m_{\rho}$ &
0.808(11) & 0.766(13) & 0.724(15) & 0.678(18) & 0.638(18) \\
32440 (99.00 \%) & $\kappa$ &
0.13491 & 0.13527 & 0.13551 & 0.13568 & 0.13581 \\
& $m_{\pi} / m_{\rho}$ &
0.8155(88) & 0.767(11) & 0.714(12) & 0.665(15) & 0.624(19) \\
32604 (99.50 \%) & $\kappa$ &
0.13443 & 0.13493 & 0.13519 & 0.13537 & 0.13551 \\
& $m_{\pi} / m_{\rho}$ &
0.8114(70) & 0.7450(92) & 0.690(12) & 0.635(16) & 0.603(18) \\
32735 (99.90 \%) & $\kappa$ &
0.13330 & 0.13391 & 0.13426 & 0.13450 & 0.13468 \\
& $m_{\pi} / m_{\rho}$ &
0.8023(42) & 0.7467(57) & 0.7009(73) & 0.6608(91) & 0.623(11) \\
32760 (99.98 \%) & $\kappa$ &
0.13268 & 0.13341 & 0.13384 & 0.13414 & 0.13436 \\
& $m_{\pi} / m_{\rho}$ &
0.8133(34) & 0.7591(48) & 0.7131(64) & 0.6718(82) & 0.636(10) \\
$8L_s^3$ (100 \%) & $\kappa$ &
0.13254 & 0.13319 & 0.13362 & 0.13394 & 0.13418 \\
& $m_{\pi} / m_{\rho}$ &
0.8071(32) & 0.7582(44) & 0.7134(59) & 0.6710(77) & 0.6332(99) \\
\hline 
\end{tabular}
\label{table:high-hopping}
\end{table*}

\end{appendices}

\bibliography{quark.bib}


\begin{thebibliography}{54}
\ifx \bisbn   \undefined \def \bisbn  #1{ISBN #1}\fi
\ifx \binits  \undefined \def \binits#1{#1}\fi
\ifx \bauthor  \undefined \def \bauthor#1{#1}\fi
\ifx \batitle  \undefined \def \batitle#1{#1}\fi
\ifx \bjtitle  \undefined \def \bjtitle#1{#1}\fi
\ifx \bvolume  \undefined \def \bvolume#1{\textbf{#1}}\fi
\ifx \byear  \undefined \def \byear#1{#1}\fi
\ifx \bissue  \undefined \def \bissue#1{#1}\fi
\ifx \bfpage  \undefined \def \bfpage#1{#1}\fi
\ifx \blpage  \undefined \def \blpage #1{#1}\fi
\ifx \burl  \undefined \def \burl#1{\textsf{#1}}\fi
\ifx \doiurl  \undefined \def \doiurl#1{\url{https://doi.org/#1}}\fi
\ifx \betal  \undefined \def \betal{\textit{et al.}}\fi
\ifx \binstitute  \undefined \def \binstitute#1{#1}\fi
\ifx \binstitutionaled  \undefined \def \binstitutionaled#1{#1}\fi
\ifx \bctitle  \undefined \def \bctitle#1{#1}\fi
\ifx \beditor  \undefined \def \beditor#1{#1}\fi
\ifx \bpublisher  \undefined \def \bpublisher#1{#1}\fi
\ifx \bbtitle  \undefined \def \bbtitle#1{#1}\fi
\ifx \bedition  \undefined \def \bedition#1{#1}\fi
\ifx \bseriesno  \undefined \def \bseriesno#1{#1}\fi
\ifx \blocation  \undefined \def \blocation#1{#1}\fi
\ifx \bsertitle  \undefined \def \bsertitle#1{#1}\fi
\ifx \bsnm \undefined \def \bsnm#1{#1}\fi
\ifx \bsuffix \undefined \def \bsuffix#1{#1}\fi
\ifx \bparticle \undefined \def \bparticle#1{#1}\fi
\ifx \barticle \undefined \def \barticle#1{#1}\fi
\bibcommenthead
\ifx \bconfdate \undefined \def \bconfdate #1{#1}\fi
\ifx \botherref \undefined \def \botherref #1{#1}\fi
\ifx \url \undefined \def \url#1{\textsf{#1}}\fi
\ifx \bchapter \undefined \def \bchapter#1{#1}\fi
\ifx \bbook \undefined \def \bbook#1{#1}\fi
\ifx \bcomment \undefined \def \bcomment#1{#1}\fi
\ifx \oauthor \undefined \def \oauthor#1{#1}\fi
\ifx \citeauthoryear \undefined \def \citeauthoryear#1{#1}\fi
\ifx \endbibitem  \undefined \def \endbibitem {}\fi
\ifx \bconflocation  \undefined \def \bconflocation#1{#1}\fi
\ifx \arxivurl  \undefined \def \arxivurl#1{\textsf{#1}}\fi
\csname PreBibitemsHook\endcsname

\bibitem[\protect\citeauthoryear{Politzer}{1973}]{Politzer:1973fx}
\begin{barticle}
\bauthor{\bsnm{Politzer}, \binits{H.D.}}:
\batitle{{Reliable Perturbative Results for Strong Interactions?}}
\bjtitle{Phys. Rev. Lett.}
\bvolume{30},
\bfpage{1346}--\blpage{1349}
(\byear{1973})
\doiurl{10.1103/PhysRevLett.30.1346}
\end{barticle}
\endbibitem

\bibitem[\protect\citeauthoryear{Gross and Wilczek}{1973}]{Gross:1973id}
\begin{barticle}
\bauthor{\bsnm{Gross}, \binits{D.J.}},
\bauthor{\bsnm{Wilczek}, \binits{F.}}:
\batitle{{Ultraviolet Behavior of Nonabelian Gauge Theories}}.
\bjtitle{Phys. Rev. Lett.}
\bvolume{30},
\bfpage{1343}--\blpage{1346}
(\byear{1973})
\doiurl{10.1103/PhysRevLett.30.1343}
\end{barticle}
\endbibitem

\bibitem[\protect\citeauthoryear{Wilson}{1974}]{Wilson:1974sk}
\begin{barticle}
\bauthor{\bsnm{Wilson}, \binits{K.G.}}:
\batitle{Confinement of quarks}.
\bjtitle{Phys. Rev. D}
\bvolume{10},
\bfpage{2445}--\blpage{2459}
(\byear{1974})
\doiurl{10.1103/PhysRevD.10.2445}
\end{barticle}
\endbibitem

\bibitem[\protect\citeauthoryear{Creutz}{1980}]{Creutz:1980zw}
\begin{barticle}
\bauthor{\bsnm{Creutz}, \binits{M.}}:
\batitle{{Monte Carlo Study of Quantized SU(2) Gauge Theory}}.
\bjtitle{Phys. Rev. D}
\bvolume{21},
\bfpage{2308}--\blpage{2315}
(\byear{1980})
\doiurl{10.1103/PhysRevD.21.2308}
\end{barticle}
\endbibitem

\bibitem[\protect\citeauthoryear{Gell-Mann}{1964}]{Gell-Mann:1964ewy}
\begin{barticle}
\bauthor{\bsnm{Gell-Mann}, \binits{M.}}:
\batitle{{A Schematic Model of Baryons and Mesons}}.
\bjtitle{Phys. Lett.}
\bvolume{8},
\bfpage{214}--\blpage{215}
(\byear{1964})
\doiurl{10.1016/S0031-9163(64)92001-3}
\end{barticle}
\endbibitem

\bibitem[\protect\citeauthoryear{Zweig}{1964a}]{Zweig:1964ruk}
\begin{botherref}
\oauthor{\bsnm{Zweig}, \binits{G.}}:
{An SU(3) model for strong interaction symmetry and its breaking. Version 1}
(1964)
\end{botherref}
\endbibitem

\bibitem[\protect\citeauthoryear{Zweig}{1964b}]{Zweig:1964jf}
\begin{bbook}
\bauthor{\bsnm{Zweig}, \binits{G.}}:
In: \beditor{\bsnm{Lichtenberg}, \binits{D.B.}},
\beditor{\bsnm{Rosen}, \binits{S.P.}} (eds.)
\bbtitle{{An SU(3) model for strong interaction symmetry and its breaking. Version 2}},
pp. \bfpage{22}--\blpage{101}
(\byear{1964})
\end{bbook}
\endbibitem

\bibitem[\protect\citeauthoryear{De~Rujula et~al.}{1975}]{DeRujula:1975qlm}
\begin{barticle}
\bauthor{\bsnm{De~Rujula}, \binits{A.}},
\bauthor{\bsnm{Georgi}, \binits{H.}},
\bauthor{\bsnm{Glashow}, \binits{S.L.}}:
\batitle{{Hadron Masses in a Gauge Theory}}.
\bjtitle{Phys. Rev. D}
\bvolume{12},
\bfpage{147}--\blpage{162}
(\byear{1975})
\doiurl{10.1103/PhysRevD.12.147}
\end{barticle}
\endbibitem

\bibitem[\protect\citeauthoryear{Oka and Yazaki}{1980}]{Oka:1980ax}
\begin{barticle}
\bauthor{\bsnm{Oka}, \binits{M.}},
\bauthor{\bsnm{Yazaki}, \binits{K.}}:
\batitle{{Nuclear Force in a Quark Model}}.
\bjtitle{Phys. Lett. B}
\bvolume{90},
\bfpage{41}--\blpage{44}
(\byear{1980})
\doiurl{10.1016/0370-2693(80)90046-5}
\end{barticle}
\endbibitem

\bibitem[\protect\citeauthoryear{Oka and Yazaki}{1981}]{Oka:1981ri}
\begin{barticle}
\bauthor{\bsnm{Oka}, \binits{M.}},
\bauthor{\bsnm{Yazaki}, \binits{K.}}:
\batitle{{Short Range Part of Baryon Baryon Interaction in a Quark Model. 1. Formulation}}.
\bjtitle{Prog. Theor. Phys.}
\bvolume{66},
\bfpage{556}--\blpage{571}
(\byear{1981})
\doiurl{10.1143/PTP.66.556}
\end{barticle}
\endbibitem

\bibitem[\protect\citeauthoryear{Godfrey and Isgur}{1985}]{Godfrey:1985xj}
\begin{barticle}
\bauthor{\bsnm{Godfrey}, \binits{S.}},
\bauthor{\bsnm{Isgur}, \binits{N.}}:
\batitle{{Mesons in a Relativized Quark Model with Chromodynamics}}.
\bjtitle{Phys. Rev. D}
\bvolume{32},
\bfpage{189}--\blpage{231}
(\byear{1985})
\doiurl{10.1103/PhysRevD.32.189}
\end{barticle}
\endbibitem

\bibitem[\protect\citeauthoryear{Capstick and Isgur}{1986}]{Capstick:1986ter}
\begin{barticle}
\bauthor{\bsnm{Capstick}, \binits{S.}},
\bauthor{\bsnm{Isgur}, \binits{N.}}:
\batitle{{Baryons in a relativized quark model with chromodynamics}}.
\bjtitle{Phys. Rev. D}
\bvolume{34}(\bissue{9}),
\bfpage{2809}--\blpage{2835}
(\byear{1986})
\doiurl{10.1103/physrevd.34.2809}
\end{barticle}
\endbibitem

\bibitem[\protect\citeauthoryear{Loring et~al.}{2001}]{Loring:2001kx}
\begin{barticle}
\bauthor{\bsnm{Loring}, \binits{U.}},
\bauthor{\bsnm{Metsch}, \binits{B.C.}},
\bauthor{\bsnm{Petry}, \binits{H.R.}}:
\batitle{{The Light baryon spectrum in a relativistic quark model with instanton induced quark forces: The Nonstrange baryon spectrum and ground states}}.
\bjtitle{Eur. Phys. J. A}
\bvolume{10},
\bfpage{395}--\blpage{446}
(\byear{2001})
\doiurl{10.1007/s100500170105}
{\href{https://arxiv.org/abs/hep-ph/0103289}{{arXiv:hep-ph/0103289}}}
\end{barticle}
\endbibitem

\bibitem[\protect\citeauthoryear{Vijande et~al.}{2005}]{Vijande:2004he}
\begin{barticle}
\bauthor{\bsnm{Vijande}, \binits{J.}},
\bauthor{\bsnm{Fernandez}, \binits{F.}},
\bauthor{\bsnm{Valcarce}, \binits{A.}}:
\batitle{{Constituent quark model study of the meson spectra}}.
\bjtitle{J. Phys. G}
\bvolume{31},
\bfpage{481}
(\byear{2005})
\doiurl{10.1088/0954-3899/31/5/017}
{\href{https://arxiv.org/abs/hep-ph/0411299}{{arXiv:hep-ph/0411299}}}
\end{barticle}
\endbibitem

\bibitem[\protect\citeauthoryear{Valcarce et~al.}{2005}]{Valcarce:2005em}
\begin{barticle}
\bauthor{\bsnm{Valcarce}, \binits{A.}},
\bauthor{\bsnm{Garcilazo}, \binits{H.}},
\bauthor{\bsnm{Fernandez}, \binits{F.}},
\bauthor{\bsnm{Gonzalez}, \binits{P.}}:
\batitle{{Quark-model study of few-baryon systems}}.
\bjtitle{Rept. Prog. Phys.}
\bvolume{68},
\bfpage{965}--\blpage{1042}
(\byear{2005})
\doiurl{10.1088/0034-4885/68/5/R01}
{\href{https://arxiv.org/abs/hep-ph/0502173}{{arXiv:hep-ph/0502173}}}
\end{barticle}
\endbibitem

\bibitem[\protect\citeauthoryear{Ebert et~al.}{2008}]{Ebert:2007nw}
\begin{barticle}
\bauthor{\bsnm{Ebert}, \binits{D.}},
\bauthor{\bsnm{Faustov}, \binits{R.N.}},
\bauthor{\bsnm{Galkin}, \binits{V.O.}}:
\batitle{{Masses of excited heavy baryons in the relativistic quark model}}.
\bjtitle{Phys. Lett. B}
\bvolume{659},
\bfpage{612}--\blpage{620}
(\byear{2008})
\doiurl{10.1016/j.physletb.2007.11.037}
{\href{https://arxiv.org/abs/0705.2957}{{arXiv:0705.2957}}}
{[hep-ph]}
\end{barticle}
\endbibitem

\bibitem[\protect\citeauthoryear{Ebert et~al.}{2009}]{Ebert:2009ub}
\begin{barticle}
\bauthor{\bsnm{Ebert}, \binits{D.}},
\bauthor{\bsnm{Faustov}, \binits{R.N.}},
\bauthor{\bsnm{Galkin}, \binits{V.O.}}:
\batitle{{Mass spectra and Regge trajectories of light mesons in the relativistic quark model}}.
\bjtitle{Phys. Rev. D}
\bvolume{79},
\bfpage{114029}
(\byear{2009})
\doiurl{10.1103/PhysRevD.79.114029}
{\href{https://arxiv.org/abs/0903.5183}{{arXiv:0903.5183}}}
{[hep-ph]}
\end{barticle}
\endbibitem

\bibitem[\protect\citeauthoryear{Vijande et~al.}{2009}]{Vijande:2009kj}
\begin{barticle}
\bauthor{\bsnm{Vijande}, \binits{J.}},
\bauthor{\bsnm{Valcarce}, \binits{A.}},
\bauthor{\bsnm{Barnea}, \binits{N.}}:
\batitle{{Exotic meson-meson molecules and compact four--quark states}}.
\bjtitle{Phys. Rev. D}
\bvolume{79},
\bfpage{074010}
(\byear{2009})
\doiurl{10.1103/PhysRevD.79.074010}
{\href{https://arxiv.org/abs/0903.2949}{{arXiv:0903.2949}}}
{[hep-ph]}
\end{barticle}
\endbibitem

\bibitem[\protect\citeauthoryear{Liu et~al.}{2019}]{Liu:2019zoy}
\begin{barticle}
\bauthor{\bsnm{Liu}, \binits{Y.-R.}},
\bauthor{\bsnm{Chen}, \binits{H.-X.}},
\bauthor{\bsnm{Chen}, \binits{W.}},
\bauthor{\bsnm{Liu}, \binits{X.}},
\bauthor{\bsnm{Zhu}, \binits{S.-L.}}:
\batitle{{Pentaquark and Tetraquark states}}.
\bjtitle{Prog. Part. Nucl. Phys.}
\bvolume{107},
\bfpage{237}--\blpage{320}
(\byear{2019})
\doiurl{10.1016/j.ppnp.2019.04.003}
{\href{https://arxiv.org/abs/1903.11976}{{arXiv:1903.11976}}}
{[hep-ph]}
\end{barticle}
\endbibitem

\bibitem[\protect\citeauthoryear{Brambilla et~al.}{2020}]{Brambilla:2019esw}
\begin{barticle}
\bauthor{\bsnm{Brambilla}, \binits{N.}},
\bauthor{\bsnm{Eidelman}, \binits{S.}},
\bauthor{\bsnm{Hanhart}, \binits{C.}},
\bauthor{\bsnm{Nefediev}, \binits{A.}},
\bauthor{\bsnm{Shen}, \binits{C.-P.}},
\bauthor{\bsnm{Thomas}, \binits{C.E.}},
\bauthor{\bsnm{Vairo}, \binits{A.}},
\bauthor{\bsnm{Yuan}, \binits{C.-Z.}}:
\batitle{{The $XYZ$ states: experimental and theoretical status and perspectives}}.
\bjtitle{Phys. Rept.}
\bvolume{873},
\bfpage{1}--\blpage{154}
(\byear{2020})
\doiurl{10.1016/j.physrep.2020.05.001}
{\href{https://arxiv.org/abs/1907.07583}{{arXiv:1907.07583}}}
{[hep-ex]}
\end{barticle}
\endbibitem

\bibitem[\protect\citeauthoryear{Workman et~al.}{2022}]{ParticleDataGroup:2022pth}
\begin{barticle}
\bauthor{\bsnm{Workman}, \binits{R.L.}}, \betal:
\batitle{{Review of Particle Physics}}.
\bjtitle{PTEP}
\bvolume{2022},
\bfpage{083}--\blpage{01}
(\byear{2022})
\doiurl{10.1093/ptep/ptac097}
\end{barticle}
\endbibitem

\bibitem[\protect\citeauthoryear{Suganuma}{2023}]{Suganuma:2023mml}
\begin{bbook}
\bauthor{\bsnm{Suganuma}, \binits{H.}}:
In: \beditor{\bsnm{Tanihata}, \binits{I.}},
\beditor{\bsnm{Toki}, \binits{H.}},
\beditor{\bsnm{Kajino}, \binits{T.}} (eds.)
\bbtitle{{Quantum Chromodynamics, Quark Confinement, and Chiral Symmetry Breaking: A Bridge Between Elementary Particle Physics and Nuclear Physics}},
pp. \bfpage{1}--\blpage{48}
(\byear{2023}).
\doiurl{10.1007/978-981-15-8818-1_22-1}
\end{bbook}
\endbibitem

\bibitem[\protect\citeauthoryear{Miransky}{1983}]{Miransky:1983vj}
\begin{barticle}
\bauthor{\bsnm{Miransky}, \binits{V.A.}}:
\batitle{{ON THE QUARK EFFECTIVE MASS IN QCD. (IN RUSSIAN)}}.
\bjtitle{Sov. J. Nucl. Phys.}
\bvolume{38},
\bfpage{280}
(\byear{1983})
\end{barticle}
\endbibitem

\bibitem[\protect\citeauthoryear{Higashijima}{1984}]{Higashijima:1983gx}
\begin{barticle}
\bauthor{\bsnm{Higashijima}, \binits{K.}}:
\batitle{{Dynamical Chiral Symmetry Breaking}}.
\bjtitle{Phys. Rev. D}
\bvolume{29},
\bfpage{1228}
(\byear{1984})
\doiurl{10.1103/PhysRevD.29.1228}
\end{barticle}
\endbibitem

\bibitem[\protect\citeauthoryear{Skullerud and Williams}{2001}]{Skullerud:2000un}
\begin{barticle}
\bauthor{\bsnm{Skullerud}, \binits{J.I.}},
\bauthor{\bsnm{Williams}, \binits{A.G.}}:
\batitle{{Quark propagator in Landau gauge}}.
\bjtitle{Phys. Rev. D}
\bvolume{63},
\bfpage{054508}
(\byear{2001})
\doiurl{10.1103/PhysRevD.63.054508}
{\href{https://arxiv.org/abs/hep-lat/0007028}{{arXiv:hep-lat/0007028}}}
\end{barticle}
\endbibitem

\bibitem[\protect\citeauthoryear{Skullerud et~al.}{2001}]{Skullerud:2001aw}
\begin{barticle}
\bauthor{\bsnm{Skullerud}, \binits{J.}},
\bauthor{\bsnm{Leinweber}, \binits{D.B.}},
\bauthor{\bsnm{Williams}, \binits{A.G.}}:
\batitle{{Nonperturbative improvement and tree level correction of the quark propagator}}.
\bjtitle{Phys. Rev. D}
\bvolume{64},
\bfpage{074508}
(\byear{2001})
\doiurl{10.1103/PhysRevD.64.074508}
{\href{https://arxiv.org/abs/hep-lat/0102013}{{arXiv:hep-lat/0102013}}}
\end{barticle}
\endbibitem

\bibitem[\protect\citeauthoryear{Bowman et~al.}{2002}]{Bowman:2002bm}
\begin{barticle}
\bauthor{\bsnm{Bowman}, \binits{P.O.}},
\bauthor{\bsnm{Heller}, \binits{U.M.}},
\bauthor{\bsnm{Williams}, \binits{A.G.}}:
\batitle{{Lattice quark propagator with staggered quarks in Landau and Laplacian gauges}}.
\bjtitle{Phys. Rev. D}
\bvolume{66},
\bfpage{014505}
(\byear{2002})
\doiurl{10.1103/PhysRevD.66.014505}
{\href{https://arxiv.org/abs/hep-lat/0203001}{{arXiv:hep-lat/0203001}}}
\end{barticle}
\endbibitem

\bibitem[\protect\citeauthoryear{Bonnet et~al.}{2002}]{Bonnet:2002ih}
\begin{barticle}
\bauthor{\bsnm{Bonnet}, \binits{F.D.R.}},
\bauthor{\bsnm{Bowman}, \binits{P.O.}},
\bauthor{\bsnm{Leinweber}, \binits{D.B.}},
\bauthor{\bsnm{Williams}, \binits{A.G.}},
\bauthor{\bsnm{Zhang}, \binits{J.-b.}}:
\batitle{{Overlap quark propagator in Landau gauge}}.
\bjtitle{Phys. Rev. D}
\bvolume{65},
\bfpage{114503}
(\byear{2002})
\doiurl{10.1103/PhysRevD.65.114503}
{\href{https://arxiv.org/abs/hep-lat/0202003}{{arXiv:hep-lat/0202003}}}
\end{barticle}
\endbibitem

\bibitem[\protect\citeauthoryear{Zhang et~al.}{2004}]{Zhang:2003faa}
\begin{barticle}
\bauthor{\bsnm{Zhang}, \binits{J.B.}},
\bauthor{\bsnm{Bowman}, \binits{P.O.}},
\bauthor{\bsnm{Leinweber}, \binits{D.B.}},
\bauthor{\bsnm{Williams}, \binits{A.G.}},
\bauthor{\bsnm{Bonnet}, \binits{F.D.R.}}:
\batitle{{Scaling behavior of the overlap quark propagator in Landau gauge}}.
\bjtitle{Phys. Rev. D}
\bvolume{70},
\bfpage{034505}
(\byear{2004})
\doiurl{10.1103/PhysRevD.70.034505}
{\href{https://arxiv.org/abs/hep-lat/0301018}{{arXiv:hep-lat/0301018}}}
\end{barticle}
\endbibitem

\bibitem[\protect\citeauthoryear{Bowman et~al.}{2004}]{Bowman:2004jm}
\begin{barticle}
\bauthor{\bsnm{Bowman}, \binits{P.O.}},
\bauthor{\bsnm{Heller}, \binits{U.M.}},
\bauthor{\bsnm{Leinweber}, \binits{D.B.}},
\bauthor{\bsnm{Parappilly}, \binits{M.B.}},
\bauthor{\bsnm{Williams}, \binits{A.G.}}:
\batitle{{Unquenched gluon propagator in Landau gauge}}.
\bjtitle{Phys. Rev. D}
\bvolume{70},
\bfpage{034509}
(\byear{2004})
\doiurl{10.1103/PhysRevD.70.034509}
{\href{https://arxiv.org/abs/hep-lat/0402032}{{arXiv:hep-lat/0402032}}}
\end{barticle}
\endbibitem

\bibitem[\protect\citeauthoryear{Szczepaniak and Swanson}{2001}]{Szczepaniak:2001rg}
\begin{barticle}
\bauthor{\bsnm{Szczepaniak}, \binits{A.P.}},
\bauthor{\bsnm{Swanson}, \binits{E.S.}}:
\batitle{{Coulomb gauge QCD, confinement, and the constituent representation}}.
\bjtitle{Phys. Rev. D}
\bvolume{65},
\bfpage{025012}
(\byear{2001})
\doiurl{10.1103/PhysRevD.65.025012}
{\href{https://arxiv.org/abs/hep-ph/0107078}{{arXiv:hep-ph/0107078}}}
\end{barticle}
\endbibitem

\bibitem[\protect\citeauthoryear{'t~Hooft}{1981}]{tHooft:1981bkw}
\begin{barticle}
\bauthor{\bsnm{Hooft}, \binits{G.}}:
\batitle{{Topology of the Gauge Condition and New Confinement Phases in Nonabelian Gauge Theories}}.
\bjtitle{Nucl. Phys. B}
\bvolume{190},
\bfpage{455}--\blpage{478}
(\byear{1981})
\doiurl{10.1016/0550-3213(81)90442-9}
\end{barticle}
\endbibitem

\bibitem[\protect\citeauthoryear{Kronfeld et~al.}{1987a}]{Kronfeld:1987vd}
\begin{barticle}
\bauthor{\bsnm{Kronfeld}, \binits{A.S.}},
\bauthor{\bsnm{Schierholz}, \binits{G.}},
\bauthor{\bsnm{Wiese}, \binits{U.J.}}:
\batitle{{Topology and Dynamics of the Confinement Mechanism}}.
\bjtitle{Nucl. Phys. B}
\bvolume{293},
\bfpage{461}--\blpage{478}
(\byear{1987})
\doiurl{10.1016/0550-3213(87)90080-0}
\end{barticle}
\endbibitem

\bibitem[\protect\citeauthoryear{Kronfeld et~al.}{1987b}]{Kronfeld:1987ri}
\begin{barticle}
\bauthor{\bsnm{Kronfeld}, \binits{A.S.}},
\bauthor{\bsnm{Laursen}, \binits{M.L.}},
\bauthor{\bsnm{Schierholz}, \binits{G.}},
\bauthor{\bsnm{Wiese}, \binits{U.J.}}:
\batitle{{Monopole Condensation and Color Confinement}}.
\bjtitle{Phys. Lett. B}
\bvolume{198},
\bfpage{516}--\blpage{520}
(\byear{1987})
\doiurl{10.1016/0370-2693(87)90910-5}
\end{barticle}
\endbibitem

\bibitem[\protect\citeauthoryear{Del~Debbio et~al.}{1997}]{DelDebbio:1996lih}
\begin{barticle}
\bauthor{\bsnm{Del~Debbio}, \binits{L.}},
\bauthor{\bsnm{Faber}, \binits{M.}},
\bauthor{\bsnm{Greensite}, \binits{J.}},
\bauthor{\bsnm{Olejnik}, \binits{S.}}:
\batitle{{Center dominance and Z(2) vortices in SU(2) lattice gauge theory}}.
\bjtitle{Phys. Rev. D}
\bvolume{55},
\bfpage{2298}--\blpage{2306}
(\byear{1997})
\doiurl{10.1103/PhysRevD.55.2298}
{\href{https://arxiv.org/abs/hep-lat/9610005}{{arXiv:hep-lat/9610005}}}
\end{barticle}
\endbibitem

\bibitem[\protect\citeauthoryear{Del~Debbio et~al.}{1998}]{DelDebbio:1998luz}
\begin{barticle}
\bauthor{\bsnm{Del~Debbio}, \binits{L.}},
\bauthor{\bsnm{Faber}, \binits{M.}},
\bauthor{\bsnm{Giedt}, \binits{J.}},
\bauthor{\bsnm{Greensite}, \binits{J.}},
\bauthor{\bsnm{Olejnik}, \binits{S.}}:
\batitle{{Detection of center vortices in the lattice Yang-Mills vacuum}}.
\bjtitle{Phys. Rev. D}
\bvolume{58},
\bfpage{094501}
(\byear{1998})
\doiurl{10.1103/PhysRevD.58.094501}
{\href{https://arxiv.org/abs/hep-lat/9801027}{{arXiv:hep-lat/9801027}}}
\end{barticle}
\endbibitem

\bibitem[\protect\citeauthoryear{Huang}{1982}]{Huang:1982ik}
\begin{bbook}
\bauthor{\bsnm{Huang}, \binits{K.}}:
\bbtitle{Quarks, Leptons and Gauge Fields},
(\byear{1982})
\end{bbook}
\endbibitem

\bibitem[\protect\citeauthoryear{Iritani and Suganuma}{2012}]{Iritani:2012bc}
\begin{barticle}
\bauthor{\bsnm{Iritani}, \binits{T.}},
\bauthor{\bsnm{Suganuma}, \binits{H.}}:
\batitle{{Lattice QCD analysis for Faddeev-Popov eigenmodes in terms of gluonic momentum components in the Coulomb gauge}}.
\bjtitle{Phys. Rev. D}
\bvolume{86},
\bfpage{074034}
(\byear{2012})
\doiurl{10.1103/PhysRevD.86.074034}
{\href{https://arxiv.org/abs/1204.6591}{{arXiv:1204.6591}}}
{[hep-lat]}
\end{barticle}
\endbibitem

\bibitem[\protect\citeauthoryear{Gribov}{1978}]{Gribov:1977wm}
\begin{barticle}
\bauthor{\bsnm{Gribov}, \binits{V.N.}}:
\batitle{{Quantization of Nonabelian Gauge Theories}}.
\bjtitle{Nucl. Phys. B}
\bvolume{139},
\bfpage{1}
(\byear{1978})
\doiurl{10.1016/0550-3213(78)90175-X}
\end{barticle}
\endbibitem

\bibitem[\protect\citeauthoryear{Zwanziger}{1997}]{Zwanziger:1995cv}
\begin{barticle}
\bauthor{\bsnm{Zwanziger}, \binits{D.}}:
\batitle{{Lattice Coulomb Hamiltonian and static color Coulomb field}}.
\bjtitle{Nucl. Phys. B}
\bvolume{485},
\bfpage{185}--\blpage{240}
(\byear{1997})
\doiurl{10.1016/S0550-3213(96)00566-4}
{\href{https://arxiv.org/abs/hep-th/9603203}{{arXiv:hep-th/9603203}}}
\end{barticle}
\endbibitem

\bibitem[\protect\citeauthoryear{Zwanziger}{1998}]{Zwanziger:1998ez}
\begin{barticle}
\bauthor{\bsnm{Zwanziger}, \binits{D.}}:
\batitle{{Renormalization in the Coulomb gauge and order parameter for confinement in QCD}}.
\bjtitle{Nucl. Phys. B}
\bvolume{518},
\bfpage{237}--\blpage{272}
(\byear{1998})
\doiurl{10.1016/S0550-3213(98)00031-5}
\end{barticle}
\endbibitem

\bibitem[\protect\citeauthoryear{Zwanziger}{2003}]{Zwanziger:2002sh}
\begin{barticle}
\bauthor{\bsnm{Zwanziger}, \binits{D.}}:
\batitle{{No confinement without Coulomb confinement}}.
\bjtitle{Phys. Rev. Lett.}
\bvolume{90},
\bfpage{102001}
(\byear{2003})
\doiurl{10.1103/PhysRevLett.90.102001}
{\href{https://arxiv.org/abs/hep-lat/0209105}{{arXiv:hep-lat/0209105}}}
\end{barticle}
\endbibitem

\bibitem[\protect\citeauthoryear{Greensite and Thorn}{2002}]{Greensite:2001nx}
\begin{barticle}
\bauthor{\bsnm{Greensite}, \binits{J.}},
\bauthor{\bsnm{Thorn}, \binits{C.B.}}:
\batitle{{Gluon chain model of the confining force}}.
\bjtitle{JHEP}
\bvolume{02},
\bfpage{014}
(\byear{2002})
\doiurl{10.1088/1126-6708/2002/02/014}
{\href{https://arxiv.org/abs/hep-ph/0112326}{{arXiv:hep-ph/0112326}}}
\end{barticle}
\endbibitem

\bibitem[\protect\citeauthoryear{'t~Hooft}{2003}]{tHooft:2002pmx}
\begin{barticle}
\bauthor{\bsnm{Hooft}, \binits{G.}}:
\batitle{{Perturbative confinement}}.
\bjtitle{Nucl. Phys. B Proc. Suppl.}
\bvolume{121},
\bfpage{333}--\blpage{340}
(\byear{2003})
\doiurl{10.1016/S0920-5632(03)01872-3}
{\href{https://arxiv.org/abs/hep-th/0207179}{{arXiv:hep-th/0207179}}}
\end{barticle}
\endbibitem

\bibitem[\protect\citeauthoryear{Rothe}{2012}]{Rothe:1992nt}
\begin{bbook}
\bauthor{\bsnm{Rothe}, \binits{H.J.}}:
\bbtitle{{Lattice Gauge Theories : An Introduction (Fourth Edition)}}
vol. \bseriesno{43}.
\bpublisher{World Scientific Publishing Company}, \blocation{???}
(\byear{2012}).
\doiurl{10.1142/8229}
\end{bbook}
\endbibitem

\bibitem[\protect\citeauthoryear{Iritani and Suganuma}{2011}]{Iritani:2011hve}
\begin{barticle}
\bauthor{\bsnm{Iritani}, \binits{T.}},
\bauthor{\bsnm{Suganuma}, \binits{H.}}:
\batitle{{Instantaneous Interquark Potential in Generalized Landau Gauge in SU(3) Lattice QCD: A Linkage between the Landau and the Coulomb Gauges}}.
\bjtitle{Phys. Rev. D}
\bvolume{83},
\bfpage{054502}
(\byear{2011})
\doiurl{10.1103/PhysRevD.83.054502}
{\href{https://arxiv.org/abs/1102.0920}{{arXiv:1102.0920}}}
{[hep-lat]}
\end{barticle}
\endbibitem

\bibitem[\protect\citeauthoryear{Luscher et~al.}{1997}]{Luscher:1996ug}
\begin{barticle}
\bauthor{\bsnm{Luscher}, \binits{M.}},
\bauthor{\bsnm{Sint}, \binits{S.}},
\bauthor{\bsnm{Sommer}, \binits{R.}},
\bauthor{\bsnm{Weisz}, \binits{P.}},
\bauthor{\bsnm{Wolff}, \binits{U.}}:
\batitle{{Nonperturbative O(a) improvement of lattice QCD}}.
\bjtitle{Nucl. Phys. B}
\bvolume{491},
\bfpage{323}--\blpage{343}
(\byear{1997})
\doiurl{10.1016/S0550-3213(97)00080-1}
{\href{https://arxiv.org/abs/hep-lat/9609035}{{arXiv:hep-lat/9609035}}}
\end{barticle}
\endbibitem

\bibitem[\protect\citeauthoryear{Aoki et~al.}{2003}]{JLQCD:2002zto}
\begin{barticle}
\bauthor{\bsnm{Aoki}, \binits{S.}},
\bauthor{\bsnm{Burkhalter}, \binits{R.}},
\bauthor{\bsnm{Fukugita}, \binits{M.}},
\bauthor{\bsnm{Hashimoto}, \binits{S.}},
\bauthor{\bsnm{Ishikawa}, \binits{K.-I.}},
\bauthor{\bsnm{Ishizuka}, \binits{N.}},
\bauthor{\bsnm{Iwasaki}, \binits{Y.}},
\bauthor{\bsnm{Kanaya}, \binits{K.}},
\bauthor{\bsnm{Kaneko}, \binits{T.}},
\bauthor{\bsnm{Kuramashi}, \binits{Y.}},
\bauthor{\bsnm{Okawa}, \binits{M.}},
\bauthor{\bsnm{Onogi}, \binits{T.}},
\bauthor{\bsnm{Tsutsui}, \binits{N.}},
\bauthor{\bsnm{Ukawa}, \binits{A.}},
\bauthor{\bsnm{Yamada}, \binits{N.}},
\bauthor{\bsnm{Yoshi\'e}, \binits{T.}}:
\batitle{{Light hadron spectroscopy with two flavors of O(a) improved dynamical quarks}}.
\bjtitle{Phys. Rev. D}
\bvolume{68},
\bfpage{054502}
(\byear{2003})
\doiurl{10.1103/PhysRevD.68.054502}
{\href{https://arxiv.org/abs/hep-lat/0212039}{{arXiv:hep-lat/0212039}}}
\end{barticle}
\endbibitem

\bibitem[\protect\citeauthoryear{Ishikawa et~al.}{1982}]{Ishikawa:1982tb}
\begin{barticle}
\bauthor{\bsnm{Ishikawa}, \binits{K.}},
\bauthor{\bsnm{Teper}, \binits{M.}},
\bauthor{\bsnm{Schierholz}, \binits{G.}}:
\batitle{{The Glueball Mass Spectrum in {QCD}: First Results of a Lattice Monte Carlo Calculation}}.
\bjtitle{Phys. Lett. B}
\bvolume{110},
\bfpage{399}--\blpage{405}
(\byear{1982})
\doiurl{10.1016/0370-2693(82)91281-3}
\end{barticle}
\endbibitem

\bibitem[\protect\citeauthoryear{Hernandez et~al.}{2005}]{Hernandez:2005:SSF}
\begin{barticle}
\bauthor{\bsnm{Hernandez}, \binits{V.}},
\bauthor{\bsnm{Roman}, \binits{J.E.}},
\bauthor{\bsnm{Vidal}, \binits{V.}}:
\batitle{{SLEPc}: A scalable and flexible toolkit for the solution of eigenvalue problems}.
\bjtitle{{ACM} Trans. Math. Software}
\bvolume{31}(\bissue{3}),
\bfpage{351}--\blpage{362}
(\byear{2005})
\end{barticle}
\endbibitem

\bibitem[\protect\citeauthoryear{Yamamoto and Suganuma}{2008}]{Yamamoto:2008am}
\begin{barticle}
\bauthor{\bsnm{Yamamoto}, \binits{A.}},
\bauthor{\bsnm{Suganuma}, \binits{H.}}:
\batitle{{Lattice analysis for the energy scale of QCD phenomena}}.
\bjtitle{Phys. Rev. Lett.}
\bvolume{101},
\bfpage{241601}
(\byear{2008})
\doiurl{10.1103/PhysRevLett.101.241601}
{\href{https://arxiv.org/abs/0808.1120}{{arXiv:0808.1120}}}
{[hep-lat]}
\end{barticle}
\endbibitem

\bibitem[\protect\citeauthoryear{Yamamoto and Suganuma}{2009}]{Yamamoto:2008ze}
\begin{barticle}
\bauthor{\bsnm{Yamamoto}, \binits{A.}},
\bauthor{\bsnm{Suganuma}, \binits{H.}}:
\batitle{{Relevant energy scale of color confinement from lattice QCD}}.
\bjtitle{Phys. Rev. D}
\bvolume{79},
\bfpage{054504}
(\byear{2009})
\doiurl{10.1103/PhysRevD.79.054504}
{\href{https://arxiv.org/abs/0811.3845}{{arXiv:0811.3845}}}
{[hep-lat]}
\end{barticle}
\endbibitem

\bibitem[\protect\citeauthoryear{Teper}{1987}]{Teper:1987wt}
\begin{barticle}
\bauthor{\bsnm{Teper}, \binits{M.}}:
\batitle{{An Improved Method for Lattice Glueball Calculations}}.
\bjtitle{Phys. Lett. B}
\bvolume{183},
\bfpage{345}
(\byear{1987})
\doiurl{10.1016/0370-2693(87)90976-2}
\end{barticle}
\endbibitem

\bibitem[\protect\citeauthoryear{Ueda et~al.}{2014}]{Ueda:2014rya}
\begin{barticle}
\bauthor{\bsnm{Ueda}, \binits{S.}},
\bauthor{\bsnm{Aoki}, \binits{S.}},
\bauthor{\bsnm{Aoyama}, \binits{T.}},
\bauthor{\bsnm{Kanaya}, \binits{K.}},
\bauthor{\bsnm{Matsufuru}, \binits{H.}},
\bauthor{\bsnm{Motoki}, \binits{S.}},
\bauthor{\bsnm{Namekawa}, \binits{Y.}},
\bauthor{\bsnm{Nemura}, \binits{H.}},
\bauthor{\bsnm{Taniguchi}, \binits{Y.}},
\bauthor{\bsnm{Ukita}, \binits{N.}}:
\batitle{{Development of an object oriented lattice QCD code 'Bridge++'}}.
\bjtitle{J. Phys. Conf. Ser.}
\bvolume{523},
\bfpage{012046}
(\byear{2014})
\doiurl{10.1088/1742-6596/523/1/012046}
\end{barticle}
\endbibitem

\end{thebibliography}
\end{document}